\documentclass[a4paper,fleqn]{cas-sc}

\usepackage[authoryear]{natbib}

\begin{document}
\let\WriteBookmarks\relax
\def\floatpagepagefraction{1}
\def\textpagefraction{.001}
\shorttitle{Eta-Virginid meteoroid stream}
\shortauthors{J. Borovi\v{c}ka et~al.}

\title [mode = title]{Eta-Virginids: another asteroidal meteoroid stream}                      



\author[1]{Ji\v{r}\'{\i} Borovi\v{c}ka}[orcid=0000-0002-5569-8982]
\cormark[1]
\ead{jiri.borovicka@asu.cas.cz}

\credit{Conceptualization, Methodology, Software, Formal analysis, Investigation, Writing - Original Draft, Visualization, Project administration, Funding acquisition}

\affiliation[1]{organization={Department of Interplanetary Matter, Astronomical Institute of the Czech Academy of Sciences},
                addressline={Fri\v{c}ova 298},
                postcode={25165}, 
                postcodesep={},                
                city={Ond\v{r}ejov},
                country={Czech Republic}}

\author[1]{Pavel Spurn\'y}
\credit{Investigation, Formal analysis, Funding acquisition}

\author[1]{Pavel Koten}
\credit{Investigation, Formal analysis}

\author[1]{Gabriel Borderes~Motta}
\credit{Formal analysis, Visualization}

\author[1]{Lenka Kotkov\'a}
\credit{Data Curation}

\author[1]{Rostislav \v{S}tork}
\credit{Resources}

\author[2]{Du\v{s}an Tomko}
\credit{Resources}

\affiliation[2]{organization={Astronomical Institute of the Slovak Academy of Sciences},
                postcode={05960}, 
                postcodesep={}, 
                city={Tatransk\'a Lomnica},
                country={Slovak Republic}}

\author[3]{Thomas Weiland}
\credit{Resources}

\affiliation[3]{organization={\"{O}sterreichischer Astronomischer Verein},
                addressline={Treumanngasse 5}, 
                postcode={1130}, 
                postcodesep={}, 
                city={Wien},
                country={Austria}}

\cortext[cor1]{Corresponding author}


\begin{abstract}
Eta-Virginids is a less-known meteor shower active in March. We investigated the meteoroids of this shower using fireball data
from the European Fireball Network supplemented by video data of faint meteors. We first derived the criteria for assigning
meteors to this shower. A fragmentation model was then applied to selected shower fireballs with good deceleration data and
light curves. Meteoroid fragmentation strengths and bulk densities were derived and compared with three other showers.
We have confirmed the four year periodicity in the activity of $\eta$-Virginids and their presence in the 3:1 mean motion resonance with Jupiter. 
The orbital period of four years was directly measured
for the fireballs. Fainter meteors showed somewhat longer periods but the shower is poor in faint meteors. No member fainter than 
magnitude +1 was observed instrumentally. 
The physical properties of meteoroids are different from cometary streams and are similar to the Geminids. 
The limiting fragmentation strength of 0.5 MPa and typical bulk density of cm-sized meteoroids of 1500 kg m$^{-3}$ suggest that the parent
body is a carbonaceous asteroid.  Besides Geminids,  $\eta$-Virginids is another stream of
asteroidal origin. Some small meteoroids have densities around 2500 kg m$^{-3}$. 
Three asteroids of the sizes between 40 -- 120 meters have been found to have similar orbits but their relation
to $\eta$-Virginids remains uncertain.
\end{abstract}



\begin{keywords}
Asteroids \sep Meteoroids \sep Meteor showers 
\end{keywords}

\maketitle

\section{Introduction}

A meteor shower occurs when the Earth passes through a meteoroid stream. Meteoroid streams are formed by meteoroids
on similar orbits having a common origin. In most cases, when the parent body of a meteoroid stream is known, it is a comet \citep{Ye_CometsIII}.
The notable exception are Geminids whose parent body is asteroid (3200) Phaethon. Another meteoroid stream connected with an asteroid,
the Daytime Sextantids connected with (155140) 2005 UD, is part of the same Phaethon-Geminid complex. The Quadrantids are related to asteroid (196256)
2003 EH$_1$ but both are part of the Machholz Interplanetary Complex which also contains comet 96P/Machholz. 
Other 21 meteor showers listed by \citet{Ye_CometsIII} are linked directly to comets.

The mechanisms how comets produce meteoroid streams are well understood. The meteoroids are ejected by gas drag during the sublimation of
ices, that means during the normal cometary activity. Large amount of meteoroids can be also released during sub-catastrophic or catastrophic break-ups
comets which comets sometimes undergo from various reasons \citep{Knight_CometsIII}. On the contrary, the formation mechanism of meteoroids within the
Phaethon-Geminid complex is still debated. One possibility is that Phaethon is an extinct comet and the meteoroids were produced by normal cometary activity 
\citep[e.g.][]{Ryabova2016}. 
Other possibilities include thermal decomposition and fracture \citep{Jewitt2010}, rotational instability \citep{Jo2024} or an unspecified catastrophic event
near perihelion \citep{Cukier2023}. Phaethon has a quite low perihelion distance (0.13 AU)
and a faint tail was observed when the asteroid is close to the Sun \citep{Jewitt2013}. The tail, however, does not consist of solid material. 
It is formed by sodium thermally desorbed from the asteroid \citep{Zhang2023}.

The nature of the parent body can be judged from the physical properties of meteoroids. It has been discovered long time ago that 
cm-sized meteoroids entering Earth's atmosphere differ enormously in their penetration ability \citep{PE}. Meteoroids coming from the asteroid
belt penetrate deep and can drop meteorites under favorable conditions. Meteoroids of cometary origin, for example members of
cometary meteor showers, disintegrate high in the atmosphere. \citet{PE} classified meteoroids into four types, I, II, IIIA, and IIIB, according to
the fireball end heights. More recently, \citet{Bor22b} introduced classes Pf-I to Pf-V according to the maximum dynamic pressure reached during
the atmospheric flight. According to both classifications, most large Geminids belong to the strongest class I and Pf-I, respectively. It was noted previously
that Geminids contain the strongest meteoroids among all meteoroid streams \citep{Spurny1993, Trigo2006, Babadzhanov2009}. 
That fact suggested an asteroidal nature of
Phaethon. However, a possibility remains that meteoroids acquired their strength by a sintering process in the vicinity of the Sun \citep{Halliday1988}.

Besides simple indicators such as Pf, meteoroid physical properties can be studied by more detailed methods. Recently, \citet{Henych2024} modeled Geminids using
the semi-empirical fragmentation model of \citet{2strengths} using genetic algorithms. They found that Geminids are stronger than the CM2 
carbonaceous meteorite Winchcombe but weaker than asteroidal meteoroids (presumably ordinary chondrites) 
and concluded that Geminids are composed of compact carbonaceous material.
Earlier, the same model was used by \citet{Tauridphys} for Taurids that originate from comet 2P/Encke and are much weaker.

In this paper, we study the less-known meteoroid stream $\eta$-Virginids. Our previous work \citep{Brcek,Bor22b} indicated that $\eta$-Virginid meteoroids
are comparable in strengths with Geminids. Also \citet{Shiba2018} noted that bright $\eta$-Virginids have lower ending heights in comparison with sporadic meteors
of similar velocities. The orbits of $\eta$-Virginids do not have such a low perihelion distance as Geminids and the meteoroids are much less affected by solar
radiation.
 Their parent body is not known. The aim of this work is to study the material properties of $\eta$-Virginids in detail using the 
semi-empirical fragmentation model and to reveal the nature of their parent body. 
Since the radiant and low-inclination orbit of this shower is not markedly distinct from many sporadic meteors
of asteroidal origin, we first need to recognize which meteors belong to $\eta$-Virginids without any doubt.

\begin{figure}
	\centering
	\includegraphics[width=8.5cm]{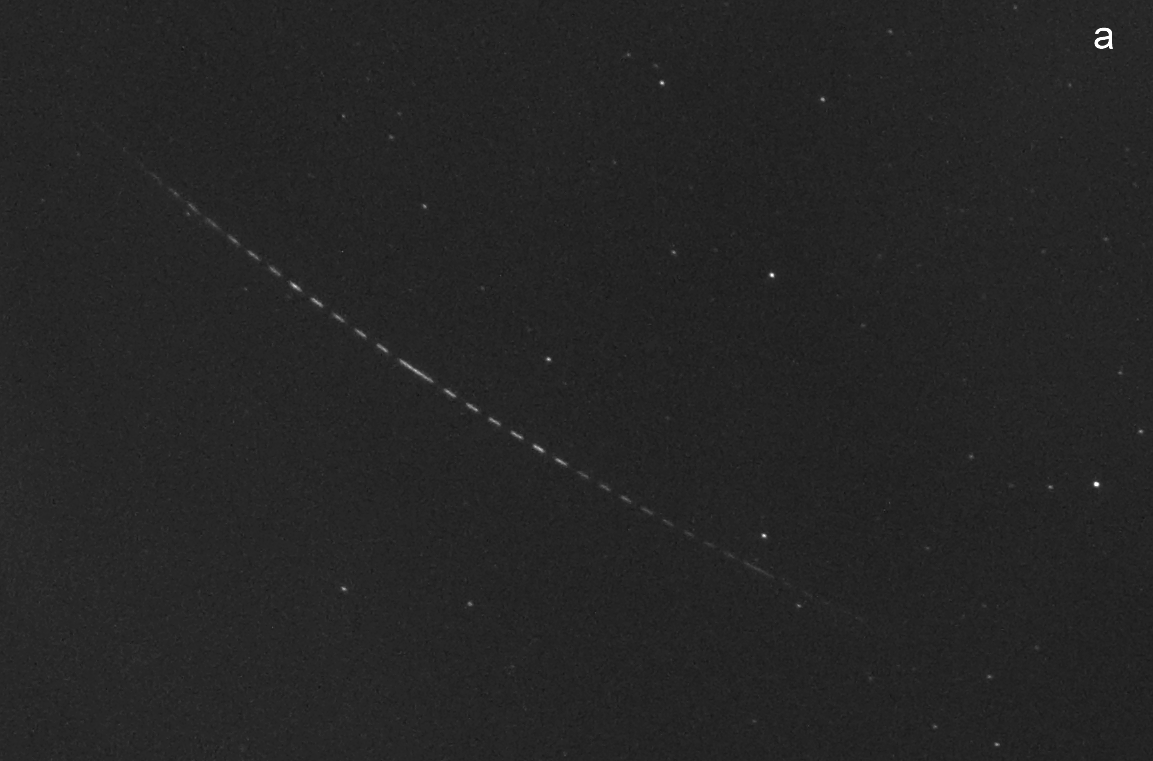}
           \includegraphics[width=7.5cm]{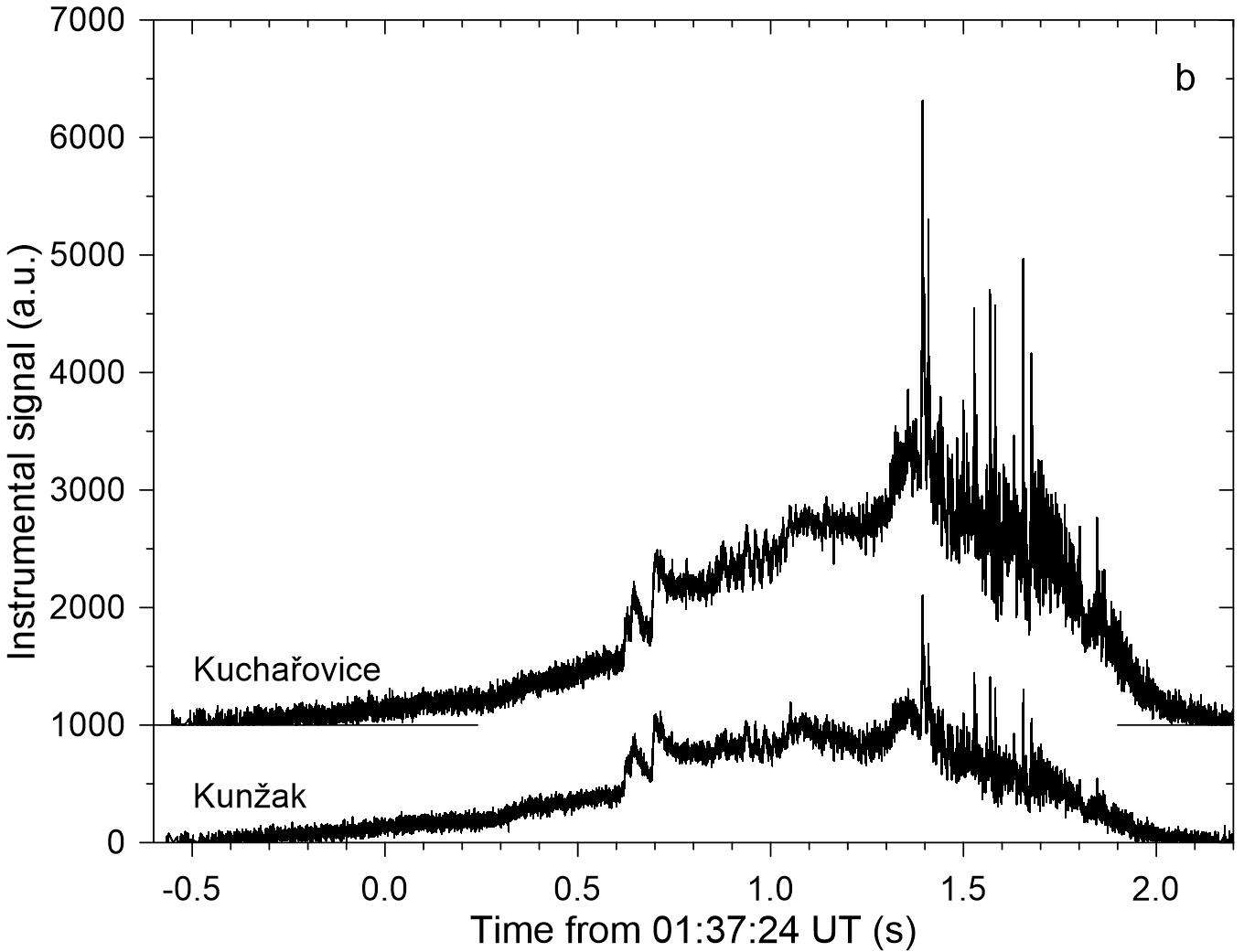}
	\caption{Eta Virginid fireball EN230321\_013723 as photographed in constellation of Hercules by DAFO at station Kun\v{z}ak (panel a) and by radiometers
	at Kun\v{z}ak and Kucha\v{r}ovice (b). 
	The sky was partly hazy that night. The breaks in the fireball image are due to the LCD shutter. One break is intentionally avoided at the beginning of each second.
	The radiometric light curve from Kucha\v{r}ovice has been offset by 1000 units for clarity.}
	\label{photo}
\end{figure}

\section{Observational data}

We primarily use fireball data obtained within the European Fireball Network (hereafter EN) \citep{Spurny17, Bor22a} in the years 2017--2025.
The main instrument is the Digital Autonomous Fireball Observatory (DAFO). DAFO takes all-sky digital photographs every clear or partly clear night.
The exposure length is 35~s and the exposure is interrupted 16 times per second by an electronic Liquid Crystal Display (LCD) shutter to
enable fireball velocity measurement (Fig.~\ref{photo}a). The limiting magnitude of DAFO under good conditions (dark clear night, meteor high in the sky)
is about $-2$. DAFOs are currently placed at 20 stations in the Czech Republic (15), Slovakia (4) and Austria (1). Combining DAFO images from at least
two stations enables us to compute the fireball trajectory, entry velocity, deceleration along the trajectory due to atmospheric drag, 
light curve with time resolution of 1/16 s, and heliocentric orbit. DAFO is further equipped with a radiometer measuring the light intensity from the whole sky
with time resolution of 1/5000 s. Radiometers are sensitive from magnitudes about $-3$ though the full time resolution can be used only for fireballs
of magnitude $-6$ and brighter. 
The radiometric curves in Fig.~\ref{photo}b demonstrate that fireball light curves can be more complex than it seems from the
photographs.

To increase the precision of the data, DAFO observations were supplemented by data from other observational systems within the EN. Spectral
version of DAFO, called SDAFO, has been installed at half of the stations plus one station in Germany. It has a lens with longer focal length but no
LCD shutter. 
Besides taking spectra of bright fireballs, 
SDAFO images can be used to improve the trajectory solutions, but not velocities. 
No $\eta$-Virginid spectrum was captured.

All stations have been also equipped with at least one Internet Protocol (IP) video camera
taking 20 or 25 frames per second.
IP cameras cover only part of the sky but are more sensitive than DAFOs and can therefore record fireballs
for a longer time. They are very useful for improving the velocity solutions
and thus also the heliocentric orbits.

In this work, we also intended to study some faint $\eta$-Virginid meteors. For that purpose, a double-station observing campaign using sensitive
image intensified video cameras was organized during three clear nights in March 2025 (March 18--20). Second generation image intensifiers 
Mullard XX1332 equipped with 1.4/50 mm  lenses providing circular field of about 50$^\circ$ were used in Ond\v{r}ejov and Kun\v{z}ak.
The videos with resolution 1280 $\times$ 960 pixels and 30 frames per second were taken with digital DMK23G445 GigE monochromatic
cameras and recorded with no compression on personal computers. The limiting magnitude for meteors was about $+5$.

The methods of data reduction are described in most detail in \citet{Bor22a}. More information about  the video system can be found in \citet{KotenTAH}.

\section{Identification of $\eta$-Virginids}

\subsection{History}

Eta Virginids are listed as an established meteor shower with number 11 and code EVI in the Meteor Shower Database of the International Astronomical Union
\citep[IAU MDC,][]{MDC}. Apart from earlier visual observations, the shower named Virginids, with an orbital period of 4 years, was mentioned by
\citet{Whipple1954} in a study of meteor orbits obtained by small photographic cameras. \citet{Jacchia1961} marked three meteors as possible
Virginids among 413 meteors photographed by more sensitive Super-Schmidt cameras, while \citet{McCrosky1961} considered five meteors from 2,539
meteors photographed by the same cameras to be Virginids. The radiants and orbits of these eight meteors varied widely. \citet{Lindblad1971}
performed a computer search for meteor showers in the catalog of \citet{McCrosky1961} and designated four meteors as Northern Virginids
and three meteors as Southern Virginids. From these seven meteors, only three were labeled as Virginids by \citet{McCrosky1961}.

In his first book, \citet{Jennbook1} named the shower $\eta$-Virginids. He included showers claimed by various authors to be active in March
in Virgo under that name. Most of those showers had a low number of observed meteors. Some were based on radar techniques. He
also listed Southern March Virginids (SVI) and Northern March Virginids (NVI) as separate showers with even longer period of activity and corresponding
to the antihelion source. The first good characterization of $\eta$-Virginids come from video observation of 56 shower members during a week
in March 2013 \citep{Jenn2016}. The shower had a narrow distribution of radiants in ecliptic latitude but extended in longitude. The mean orbital elements
had a semimajor axis $a=2.47$ AU corresponding to period 3.9 years, perihelion distance $q=0.46$ AU, and inclination $i=5.4^\circ$.

A major progress was done by \citet{Shiba2018} who analyzed Japanese video data obtained between 2007 -- 2018. 
He found a four year cycle in meteor activity with the highest $\eta$-Virginid activity being observed in the years 2009, 2013, and 2017, somewhat lower activity in the
following years 2010, 2014, and 2018 and no activity in the remaining years. He proposed existence of a meteoroid swarm in 3:1 mean motion 
resonance with Jupiter. The corresponding orbital period is 3.95 years. The orbital elements of individual meteors, however, exhibited significant scatter.
The mean orbital period was only 3.58 yr, which was attributed to calculation errors. \citet{Jennbook2} confirmed the four-year periodicity in activity
though he detected the shower every year.

\begin{figure}
	\centering
	\includegraphics[width=7.75cm]{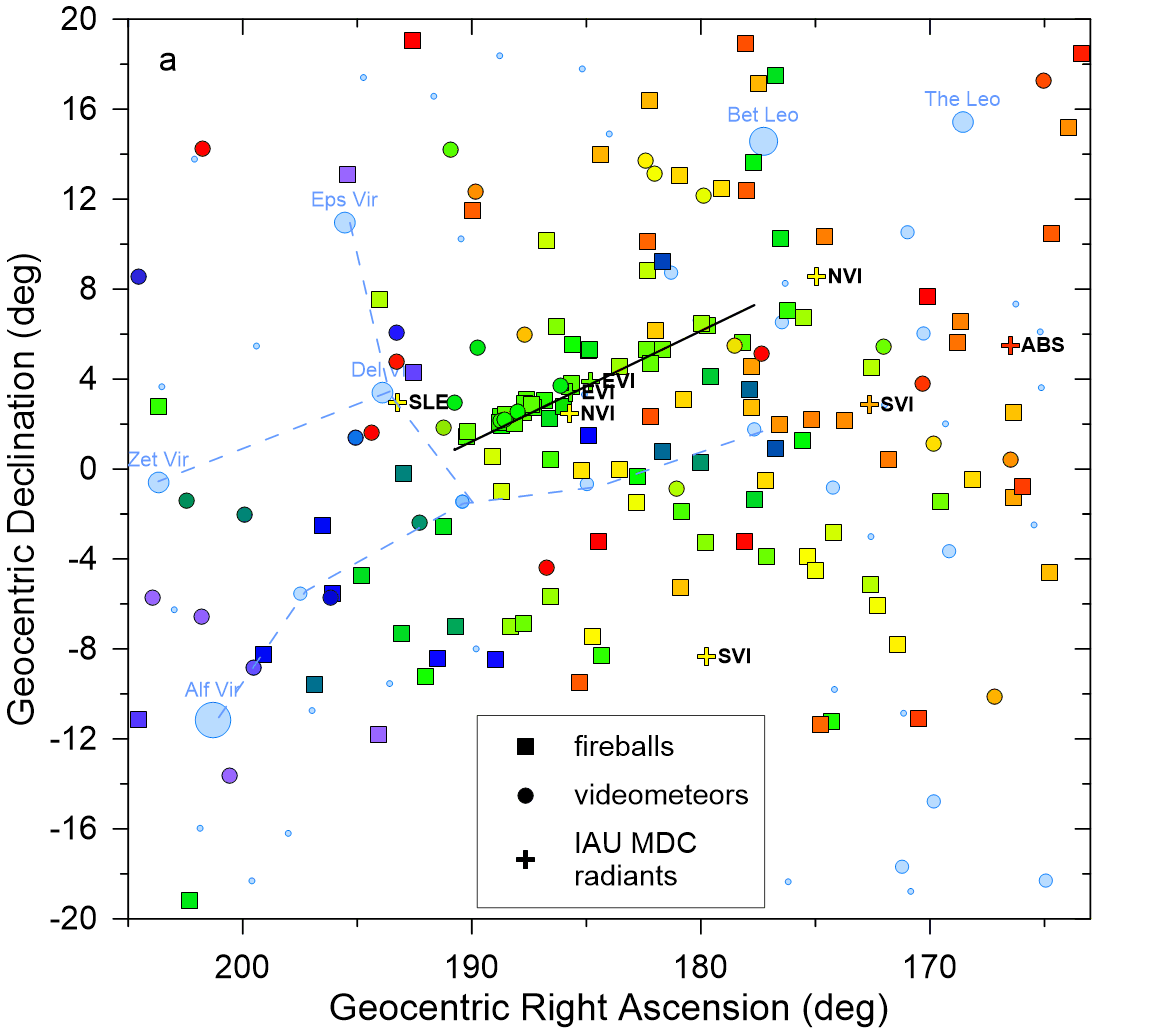}
	\includegraphics[width=8.5cm]{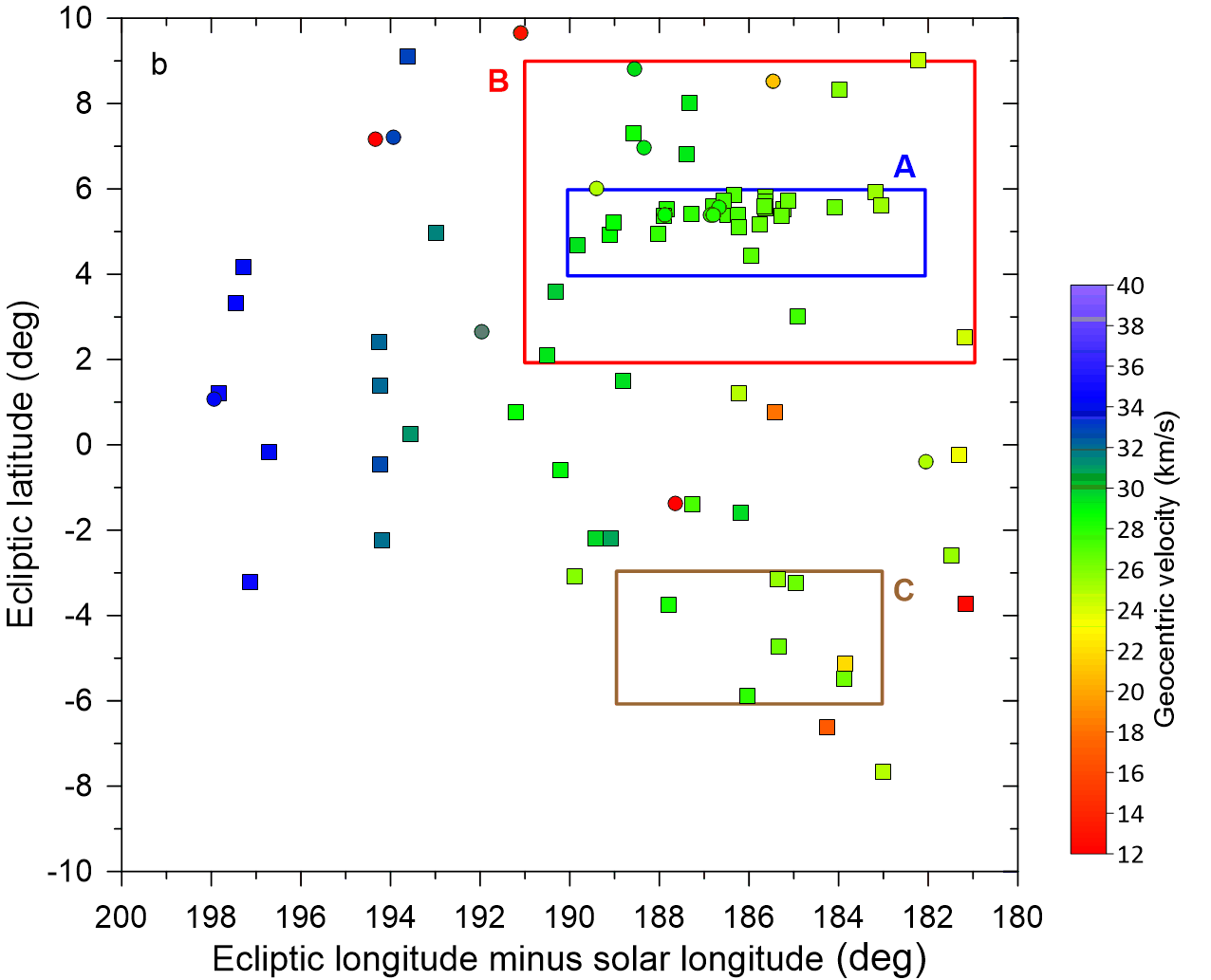}
	\caption{Positions of geocentric radiants located in the Virgo region in equatorial coordinates (panel a) and in Sun-centered ecliptical coordinates (b). 
	The radiants of all fireballs observed by the EN during the month of March in 2017--2025 are plotted as color squares. 
	The radiants of video meteors observed during three nights in March 2025
	are plotted as color circles. The radiants of meteor showers active in March according to the IAU MDC are plotted as crosses in panel a.
(NVI = Northern March Virginids, SVI = Southern March Virginids, EVI =$\eta$-Virginids, SLE = $\sigma$-Leonids, ABS = April $\beta$-Sextantids)
	Some showers have multiple solutions in the database and thus more possible radiants.
	Geocentric velocities are color coded. Stars are plotted as light blue circles. The black line marks the concentration of radiants.
	Three regions of interest are marked by rectangles in panel b:
	radiant concentration (A), wider area around the concentration (B) and possible southern branch (C).}
	\label{radiants}
\end{figure}

\subsection{EN data}

Eta Virginids obviously overlap with the antihelion source of sporadic meteors. We used EN data to develop strict criteria to assign meteors to
the shower. Figure~\ref{radiants}a shows geocentric radiants and velocities of all fireballs observed between March 1--31 of all studied years
and having radiants in the Virgo region.
The only concentration of radiants is along the marked line and overlaps with the cataloged radiant of  $\eta$-Virginids according to two sources.
One cataloged radiant of Northern March Virginids is also close but the velocity is different. We therefore definitely detected $\eta$-Virginids in our sample.
To study their physical properties, we must distinguish them from sporadic fireballs with similar radiants.

The radiants are plotted in Fig.~\ref{radiants}b in Sun-centered ecliptical coordinates, namely ecliptical longitude minus solar longitude ($\lambda - \lambda_\odot$)
and ecliptical latitude ($\beta$).  The region of radiant concentration (marked A) is narrow in $\beta$ and prolonged in $\lambda - \lambda_\odot$,
as noted already by \citet{Jenn2016}. We will investigate further if meteors from a wider region (marked B) may also belong to $\eta$-Virginids.
In addition, since \citet{Shiba2018} and \citet{Brcek} mentioned a possible southern branch, we also investigated a region on the opposite side 
of the ecliptic (marked C). Some meteors in these additional regions have obviously deviating velocities but several have velocities roughly consistent
with $\eta$-Virginids.

\begin{figure}
	\centering
	\includegraphics[width=8.5cm]{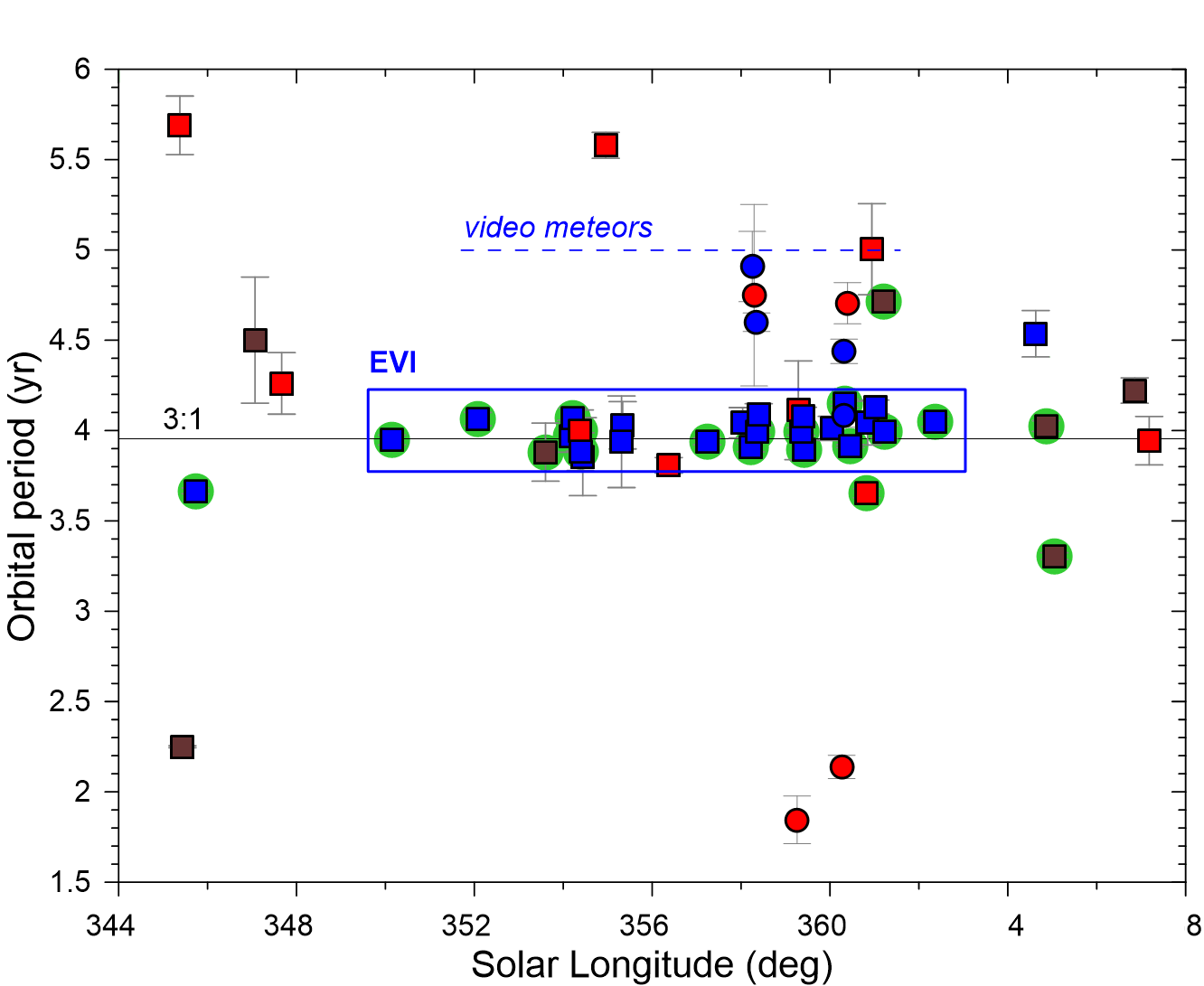}
	\caption{Orbital periods versus solar longitude of meteors with radiants in regions A (blue), B (red), and C (brown). 
	EN fireballs are plotted as squares, video meteors as circles. 
	Video data, generally, have somewhat lower precision.
	Error bars of orbital periods are indicated (one sigma). The horizontal line corresponds
	to the 3:1 resonance with Jupiter. The blue rectangle is the region of $\eta$-Virginid fireballs. The dashed blue line indicates possible
	extension of orbital periods of $\eta$-Virginid video meteors. The fireballs which were subject to fragmentation modeling (see Sect.~\ref{modeling}) 
	are highlighted by green circle.}
	\label{periods}
\end{figure}

We further investigated semimajor axes or, equivalently, orbital periods. Orbital periods of meteors with radiants in regions A, B,
 C are plotted 
in Fig.~\ref{periods} versus solar longitude, i.e.\ the time of meteor appearance. We can see that all fireballs in region A appearing between solar longitudes 
$350^\circ$ -- $3^\circ$ (nearly March 10 -- 23) had orbital periods very close to four years. One fireball appearing in this region earlier and one
appearing later had different periods and we therefore do not consider them to belong to $\eta$-Virginids. Fireballs of regions B and C show much larger scatter
in both solar longitude and period and we also do not consider them to belong to $\eta$-Virginids, though three fireballs of region B and one of region C 
fall, probably by chance, within the right limits of solar longitude and period.

In summary, our working hypothesis is that only fireballs appearing at $\lambda_\odot$ between $350^\circ$ and $3^\circ$, having radiants
with $\beta$ between $4^\circ$ and $6^\circ$ and $\lambda - \lambda_\odot$ between $182^\circ$ and $191^\circ$, and orbital periods
between 3.8 -- 4.2 years are with high probability true $\eta$-Virginids. Note that from eight marked as Virginids in the Super-Schmidt data
by \citet{Jacchia1961} and \citet{McCrosky1961}, only two (meteors no. 6798 and 6949) can be considered to be $\eta$-Virginids according to
their radiants. A similar conclusion was reached by \citet{Koseki2019}. In our previous work \citep{Brcek}, EN060319\_204604 was
incorrectly considered to be an $\eta$-Virginid.

\begin{figure}
	\centering
	\includegraphics[width=7cm]{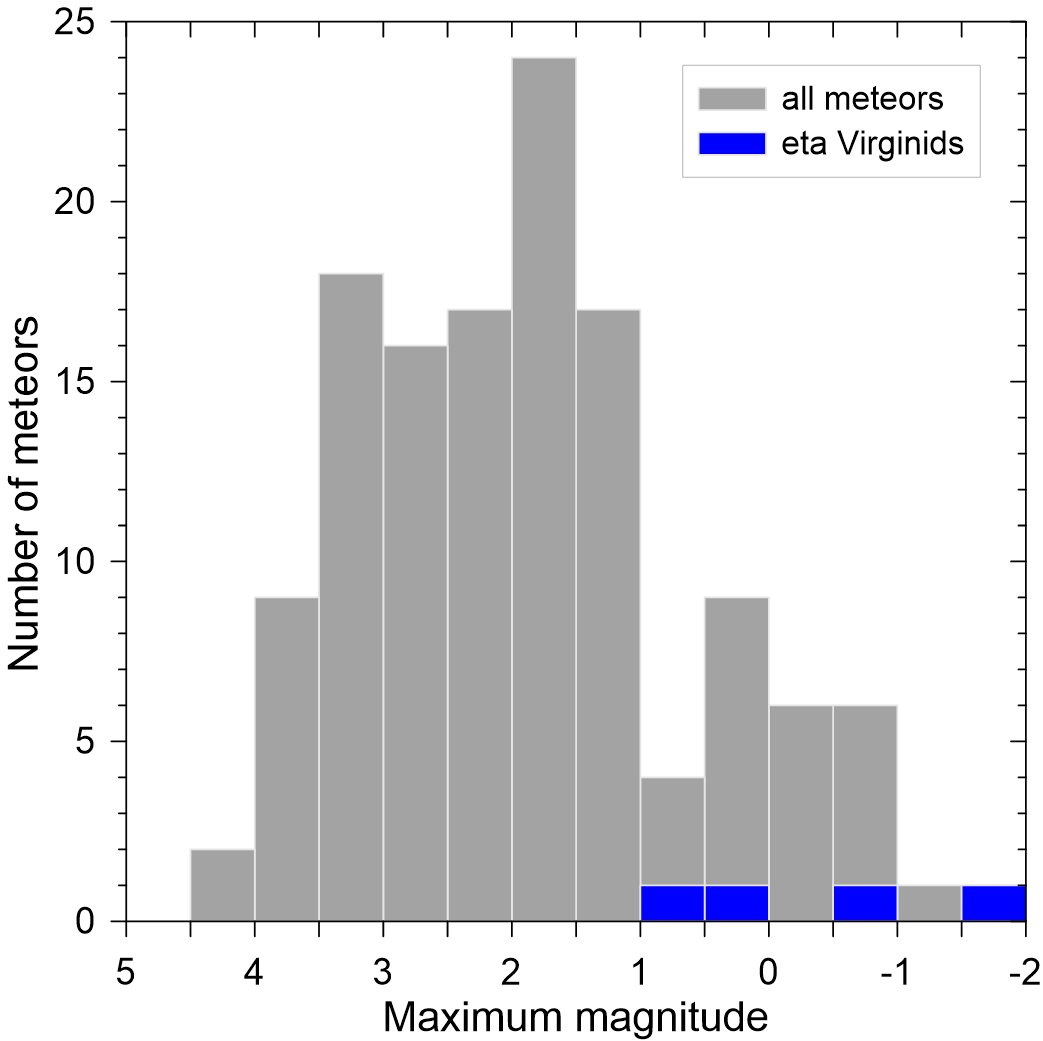}
	\caption{Magnitude distribution of all 130 meteors observed by video cameras during three nights in March 2025 and of four $\eta$-Virginids
	among them.}
	\label{hist-mag}
\end{figure}

\subsection{Video data}

In total, 130 double-station meteors were observed by the video experiment in three clear nights in March 2021 (at solar longitudes $358^\circ$
-- $0.5^\circ$). They are included as circles in Figs.~\ref{radiants} and \ref{periods}. Only four meteors had radiants in region A, four in B, and none in C.
Only one meteor had the period close to 4 years. The other three meteors from region A had periods between 4.4 -- 4.9 years. Nevertheless,
we consider them as probable $\eta$-Virginids. It is possible that small meteoroids have larger semimajor axes from some reason, for example 
the influence of radiation pressure \citep{Burns} or larger ejection velocities during stream formation. 
All four $\eta$-Virginids were brighter than magnitude +1 while large majority of other meteors were fainter than +1
(Fig.~\ref{hist-mag}). It therefore seems that $\eta$-Virginids contain very few faint meteors. The two Super-Schmidt $\eta$-Virginids (6798 and 6949) were
both brighter than zero magnitude. 
On the other hand, during the visual observations by one of us (TW) on four clear nights in 2021 (16/17 to 20/21 March),
10 of the 18 observed $\eta$-Virginids were fainter than +1 mag (down to +5). Nevertheless, the magnitude distribution was quite flat. 

\begin{figure}
	\centering
	\includegraphics[width=7cm]{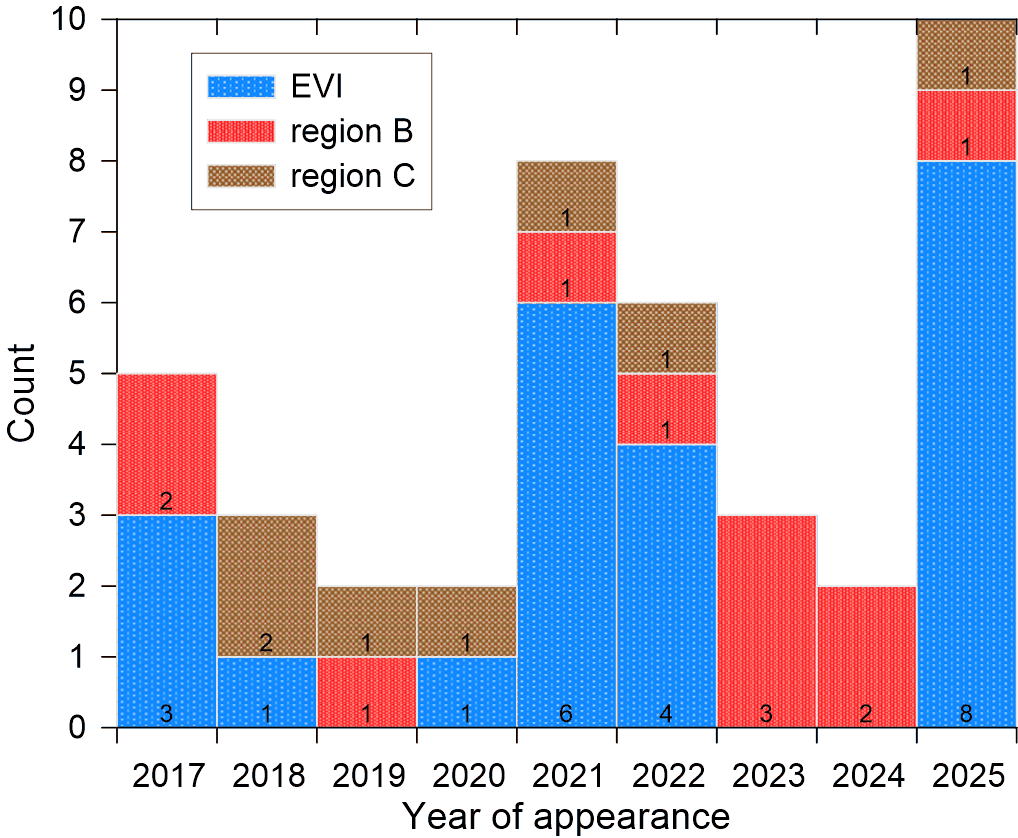}
	\caption{Number of EN fireballs observed in each year. The confirmed $\eta$-Virginids are in blue. Fireballs with radiants in region B and
	those with radiants in region A but orbital periods incompatible with $\eta$-Virginids are in red. Fireballs with radiants in region C are in brown.}
	\label{years}
\end{figure}

\section{Periodic activity}

As we have now identified the true $\eta$-Virginids, we can look in which years they were active. The distribution was not random (Fig.~\ref{years}).
A larger number of fireballs was observed in 2017, 2021, 2022, and 2025. One $\eta$-Virginid was observed in 2018 and 2020. We can therefore
confirm the observation of \citet{Shiba2018} that $\eta$-Virginids are always active in two consecutive years and no activity occurs in the next two
years. This four-year cycle is in full agreement with the 4 year orbital period we have found. The only fireball which occurred in 2020 when no activity
was expected was EN190320\_201746.
As we will see in Sect.~\ref{modeling}, that meteoroid had different physical properties than the rest of $\eta$-Virginids. It is therefore possible that it was
a random interloper and not a shower member.

Fireballs with radiants in regions B and C (see Fig.~\ref{radiants}b) are also shown in Fig.~\ref{years}. They occurred randomly and are not correlated
with $\eta$-Virginids. That fact confirms that they did not belong to the shower.

\begin{figure}
	\centering
	\includegraphics[width=8cm]{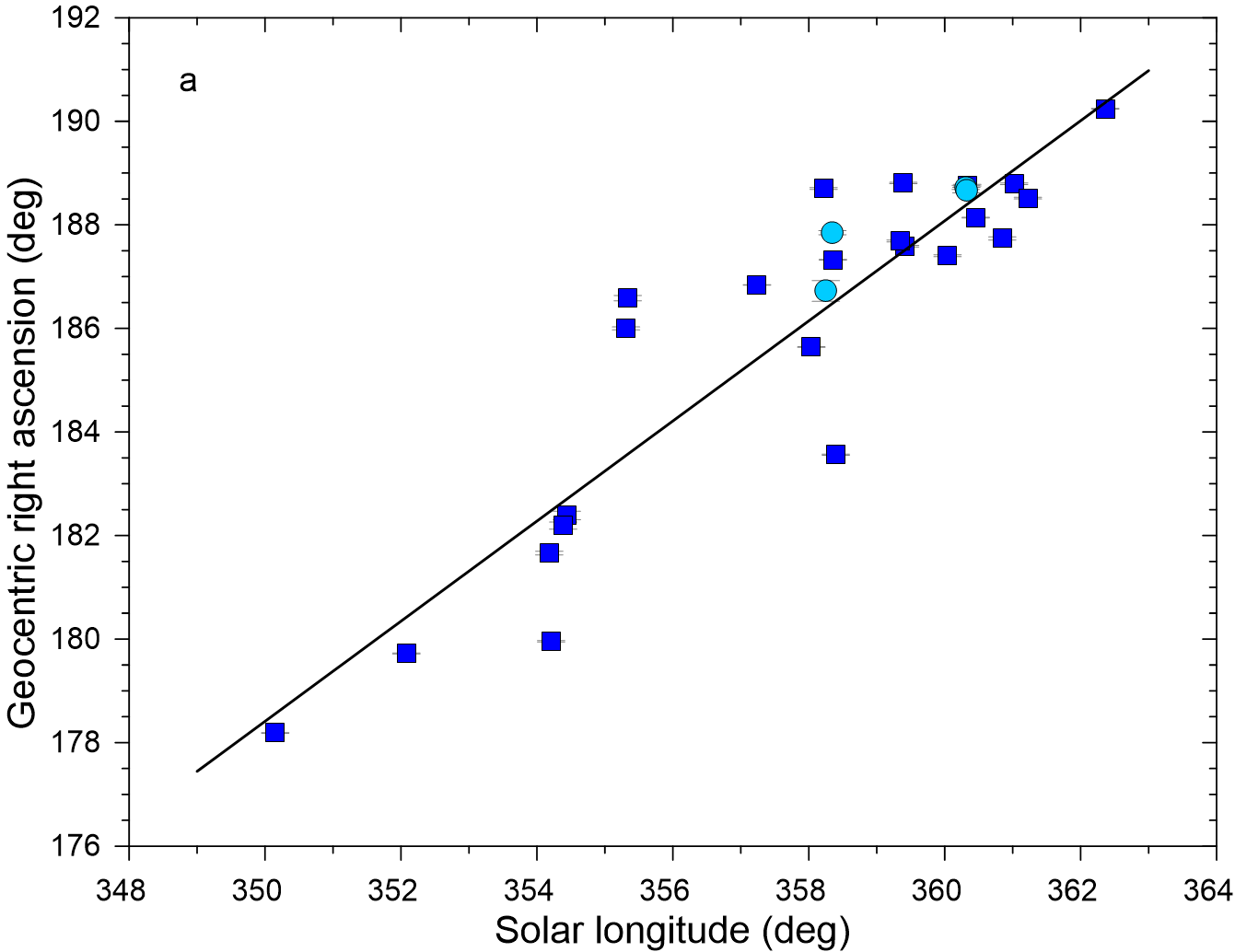}
	\includegraphics[width=8cm]{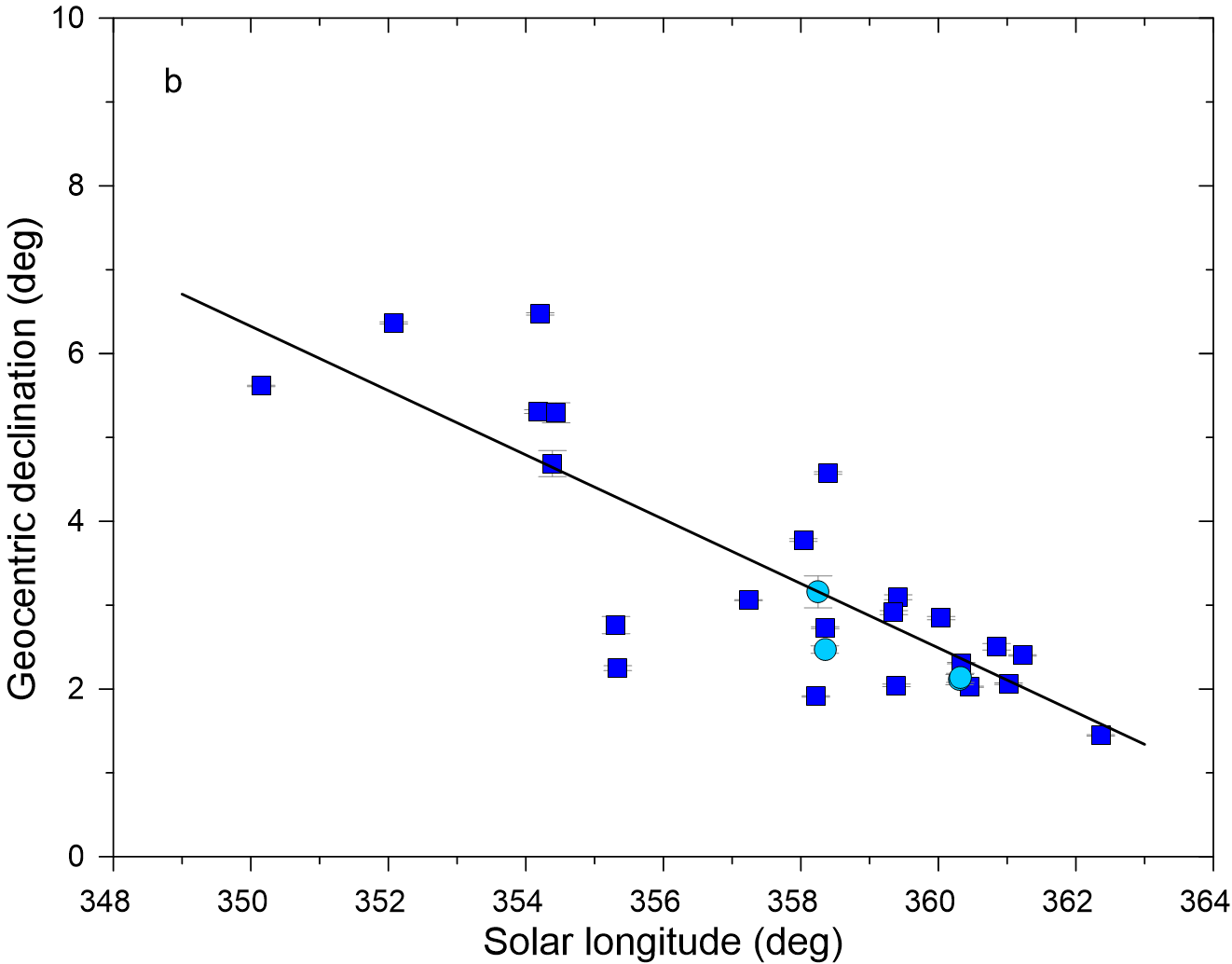}
	\caption{Right ascension (a) and declination (b) of geocentric radiant as a function of solar longitude. EN fireballs are shown as squares,
	video meteors as circles.}
	\label{radmotion}
\end{figure}

\section{Radiants and orbits}

We present in this section the mean radiants and orbits of $\eta$-Virginids based primarily on the photographic EN data. The motion of the mean
geocentric radiant with solar longitude $\lambda_\odot$ can be expressed as
\begin{eqnarray}
\alpha_{\rm g} &=& 186.1^\circ + 0.97 \cdot (\lambda_\odot - 358^\circ) \\
\delta_{\rm g} &=& +3.3^\circ - 0.38 \cdot (\lambda_\odot - 358^\circ),
\end{eqnarray}
where $\alpha_{\rm g}$ is the right ascension and $\delta_{\rm g}$ is the declination of the geocentric radiant. The shower was observed active
between solar longitudes 350$^\circ$ and 3$^\circ$ (March 10 -- 23). For solar longitudes between zero and several degrees, 360$^\circ$ must be
added to use the above equations. The radiant motion is depicted graphically in Fig.~\ref{radmotion}.

\begin{figure}
	\centering
	\includegraphics[width=8cm]{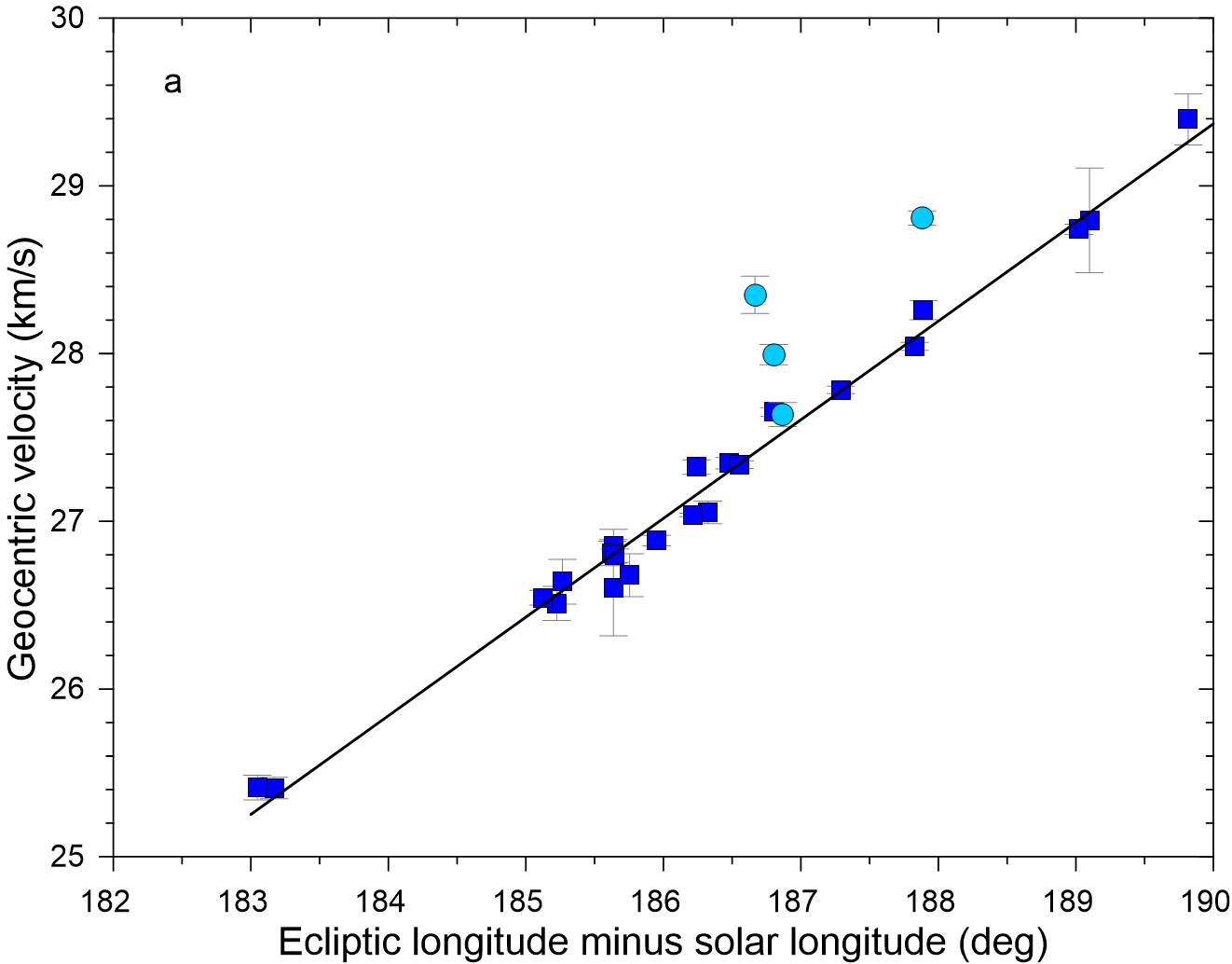}
	\includegraphics[width=8cm]{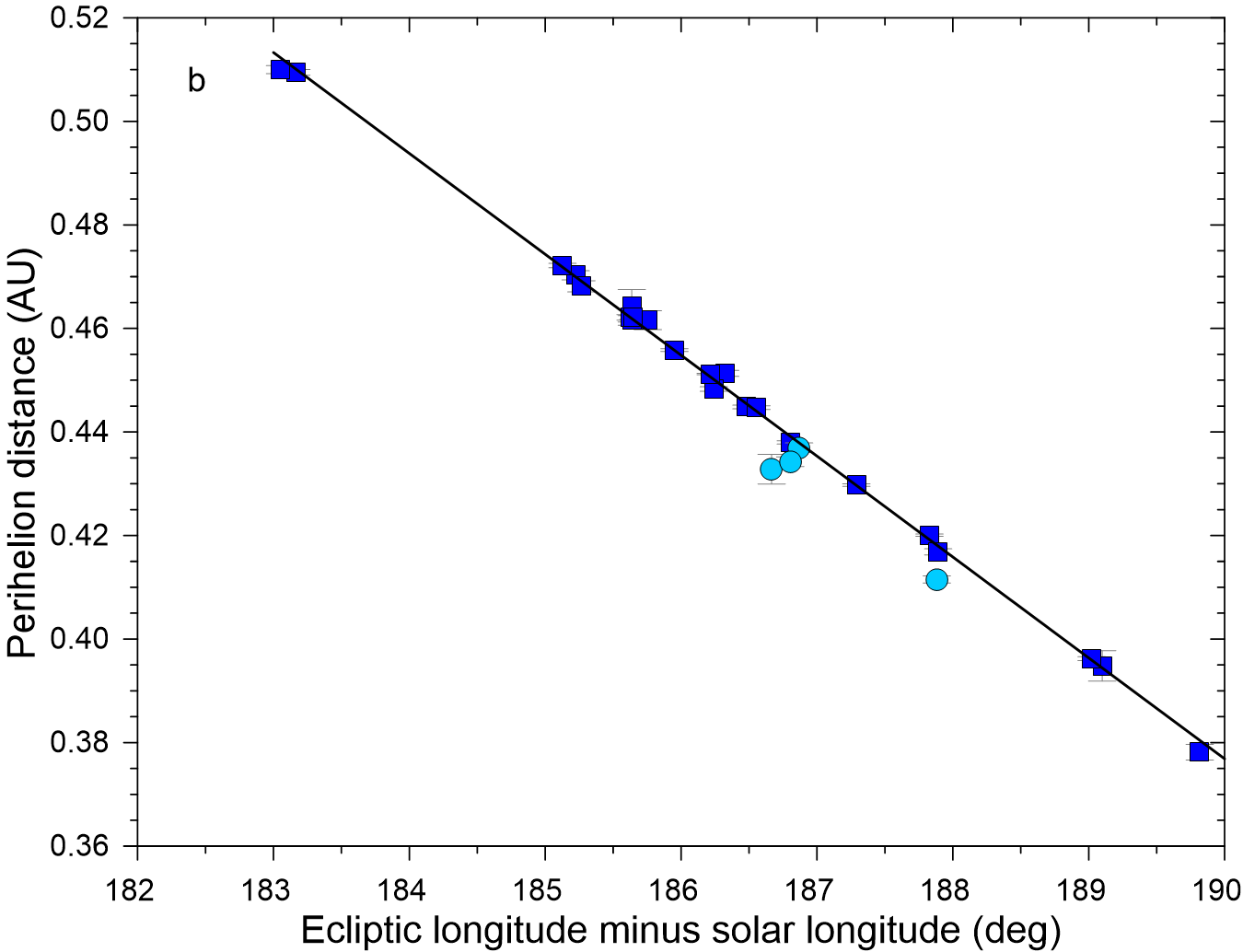}
	\caption{Geocentric velocity (a) and perihelion distance (b) as a function of Sun-centered ecliptical longitude of the radiant. EN fireballs are shown as squares,
	video meteors as circles.}
	\label{longitude}
\end{figure}

The radiant motion follows the motion of the Sun along ecliptic. As shown in Fig.~\ref{radiants}b, the spread in ecliptic latitude is low, about 1.5$^\circ$.
The spread in ecliptic longitude related to the solar longitude is 7$^\circ$. The spread is closely related to the spread of geocentric velocities and perihelion
distances, as shown in Fig.~\ref{longitude}. The fireballs with radiants more to the east from the antisolar point have larger velocities
and lower perihelion distances. The linear dependencies are very well defined for EN fireballs. Some of video meteors lie somewhat off. 
The ranges of geocentric velocities (25.5 -- 29.5 km s$^{-1}$) and perihelion distances (0.38 -- 0.51 AU) are relatively large which contrasts
with the narrow range of semimajor axes, and thus orbital periods, of the fireballs (Fig.~\ref{periods}). Similar linear dependencies
exist also for the eccentricity and the argument of perihelion. Inclinations are in a narrow range, 5 -- 6$^\circ$ for most fireballs and meteors;
only one fireball with $i=4.38^\circ$ deviates markedly.

\begin{figure}
	\centering
	\includegraphics[width=10cm]{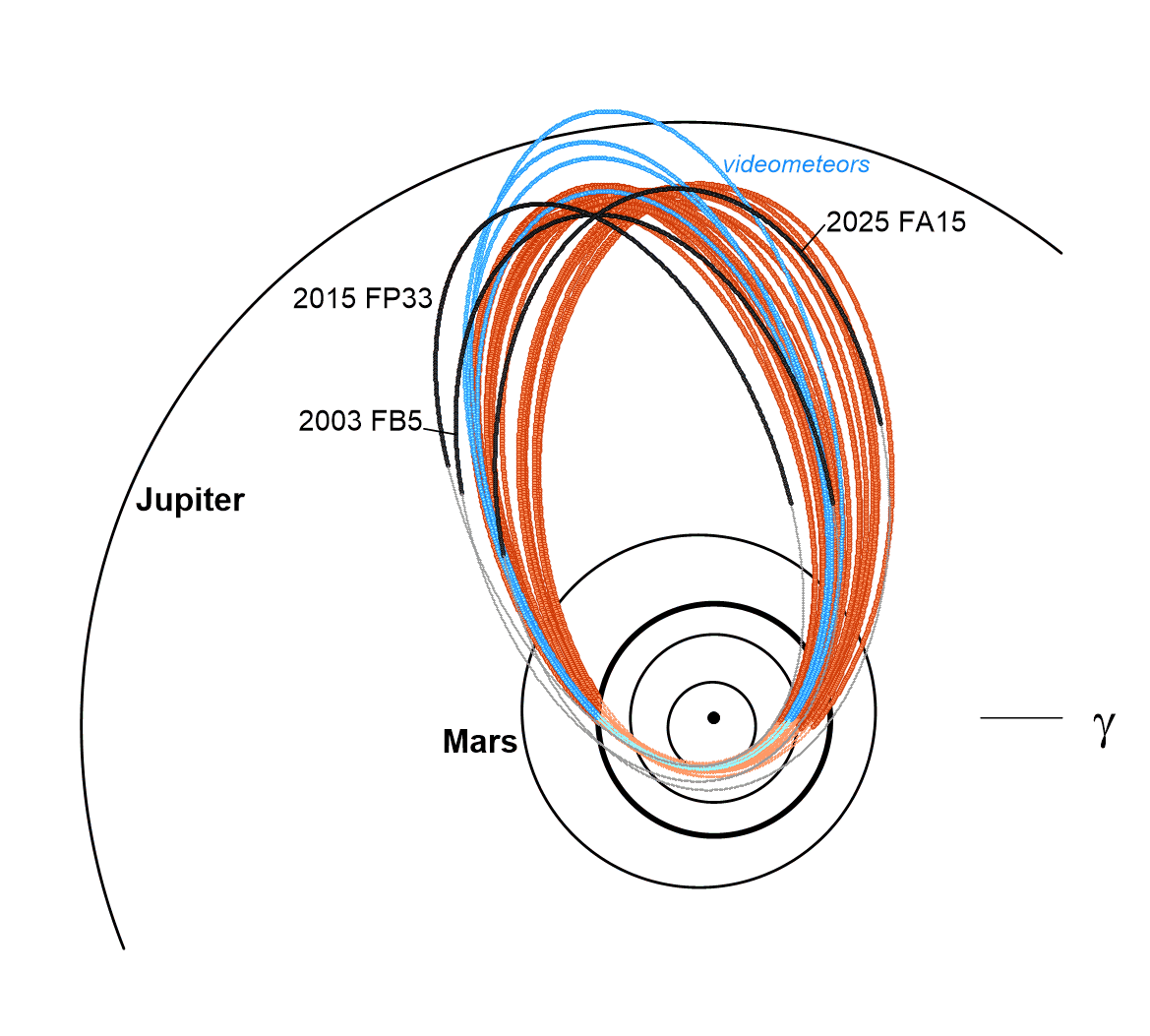}
	\caption{Orbits of all observed $\eta$-Virginids in the projection to the plane of ecliptic. EN fireballs are in red and
	video meteors in blue. The thin lines depict the parts of orbits below (to the south of) the ecliptic. The three black/gray curves show the
	orbits of three asteroids with orbits most similar to $\eta$-Virginids. The position of the Sun, the orbits of planets from Mercury to Jupiter, and
	the direction to the vernal equinox are also shown.}
	\label{orbits}
\end{figure}

\begin{table*}[width=13.5cm]
\caption{Mean orbital elements of 23 $\eta$-Virginid fireballs (J2000.0)}
\label{tab:orbit}
\begin{tabular}{lllllll}
\toprule
Semimajor & Eccentricity & Perihelion & Aphelion & Inclination & Argument & Longitude of \\
axis && distance & distance & & of perihelion & ascending node \\
AU && AU & AU & degrees & degrees & degrees \\
\midrule
2.52 & 0.82 & 0.45 & 4.6 &  5.5 & 283 & 358 \\
\bottomrule
\end{tabular}
\end{table*}

The orbits are shown graphically in Fig.~\ref{orbits}. Table~\ref{tab:orbit} contains the mean orbital elements based on the fireball data. 
The mean elements are similar to those obtained by \citet{Shiba2022}.
We must, nevertheless, keep in mind that most elements, except semimajor axis and inclination, exhibit significant natural spread, which is not
due to observational errors. The range of eccentricities is 0.80 -- 0.85, and of arguments of perihelion 276 -- 290$^\circ$. Fainter meteors observed
by video cameras show also a spread in semimajor axes.

\begin{figure}
	\centering
	\includegraphics[width=8cm]{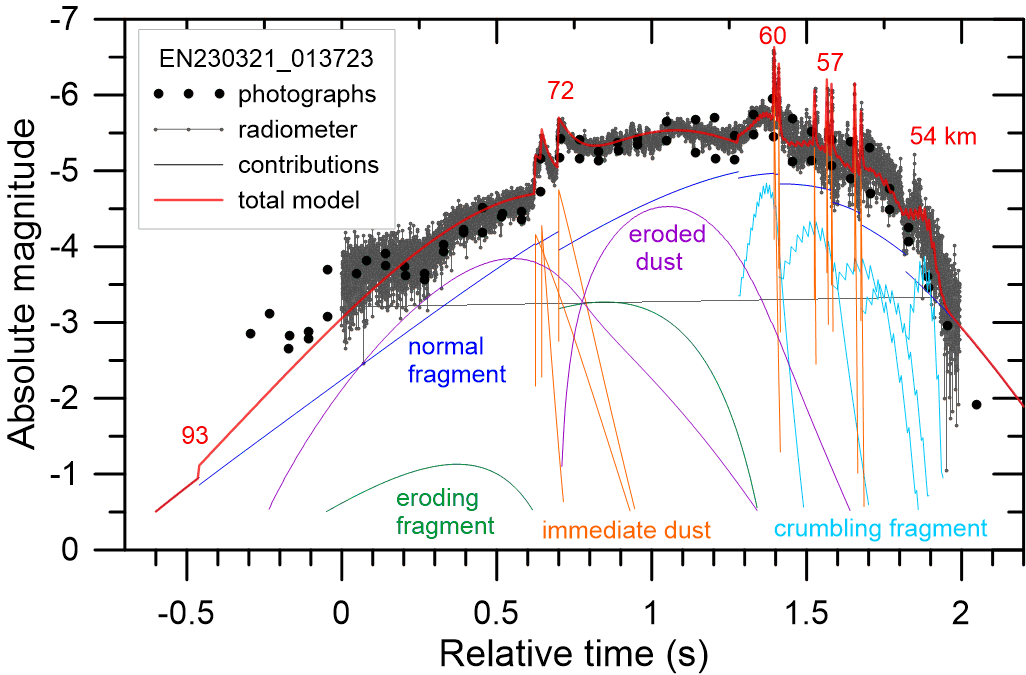}
	\includegraphics[width=8cm]{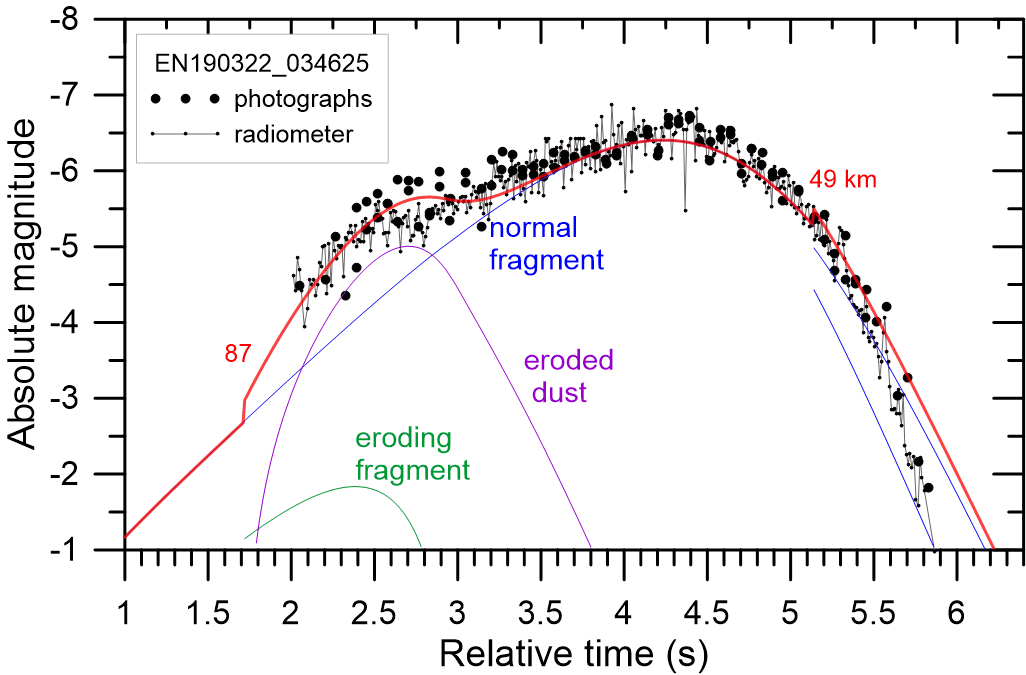}
	\caption{Observed and modeled light curves of $\eta$-Virginid fireballs EN230321\_013723 and EN190322\_034625. Individual contributors
	to the model are shown. The numbers are heights of selected fragmentation events in km. The zenith angle of EN230321\_013723 was 52$^\circ$;
	in case of  EN190322\_034625 it was 67$^\circ$.}
	\label{lc}
\end{figure}

\section{Physical properties}
\label{modeling}

\subsection{Fragmentation modeling}

The main goal of this paper is the evaluation of physical properties of $\eta$-Virginid meteoroids, namely fragmentation strengths and bulk densities.
As in our previous work on Taurids \citep{Tauridphys}, we used the semi-empirical fragmentation model. The model has been described in detail
in \citet{2strengths}. The fitting of the observed light curves and decelerations was done manually. The free parameters were the initial mass, velocity, 
and density of the meteoroid and the description of the fragmentation process during the atmospheric flight. The outputs of each fragmentation event
could be normal individual fragments, a group of fragments of identical masses, immediately released dust, an eroding fragment releasing dust gradually,
or a crumbling fragment undergoing progressive fragmentation. The ablation coefficient of all fragments and dust particles was kept constant at
$\sigma=0.005$ s$^2$km$^{-2}$, the product of the drag and shape coefficient was assumed to be $\Gamma A = 0.8$, the density of the dust particles was
2000 kg\,m$^{-3}$,
 and the mass distribution index of the dust particles was $s=2.0$. The erosion coefficient and the upper and lower mass limit
of the dust, either eroded or immediately released, was to be adjusted. The crumbling fragment is a new feature of the model describing schematically
fragments which are fragmenting in a cascade. It is assumed that each created fragment is fragmenting further after a fixed time interval into a fixed
number of fragments. The number of fragments in the cascade therefore grows geometrically and the masses of all fragments are identical at any time instant.
The free parameters are the time interval of fragmentation, the number of daughter fragments, and the mass limit at which the fragmentation stops.
The used atmospheric density and luminous efficiency model is described in \citet{2strengths}.

As an example, Fig.~\ref{lc} shows the light curve of fireball EN230321\_013723 which was imaged in Fig.~\ref{photo}. While the light curve
looks smooth on the photograph, the radiometer revealed a number of short flares with the amplitude of about one magnitude. They could be well
modeled by immediate dust releases. In overall, the fragmentation was quite complex and both eroding fragments and crumbling fragments were needed
to explain the light curve. On the other hand, another fireball displayed in Fig.~\ref{lc}, EN190322\_034625, had a really smooth light curve with
only a hump at the beginning, explained by erosion. There was also a fragment splitting near the end at the height of 49 km which was
not demonstrated on the light curve but was needed in the model to explain the increase of deceleration at the end. 

In both fireballs, a fragmentation before the 
beginning of our instrumental records,
at the heights around 90 km, was needed to explain the shape of the light curve at the beginning.
The created eroding fragment contained 10 -- 15\% of the original mass. It may represent an incoherent surface layer of the meteoroid. 
Such effect was observed in a majority, but not all, $\eta$-Virginids.

Among 23 confirmed $\eta$-Virginid fireballs, 14 fireballs had good enough data for fragmentation modeling. In addition, we modeled 
for comparison seven fireballs which were excluded from the $\eta$-Virginid list on the basis of their radiants or orbital periods. 
Four of them have radiants in the southern region C.  The modeled fireballs are highlighted in Fig.~\ref{periods}.
To evaluate the strength of the meteoroids, we computed the dynamic pressures, $p$, at the time of fragmentation events. The pressure was
computed as $p=\rho v^2$, where $\rho$ is atmospheric density and $v$ is fireball velocity. As in many similar studies, the role of the 
drag coefficient $\Gamma$ was ignored, since its actual value is not known but it is of the order of unity.

\begin{figure}
	\centering
	\includegraphics[width=10cm]{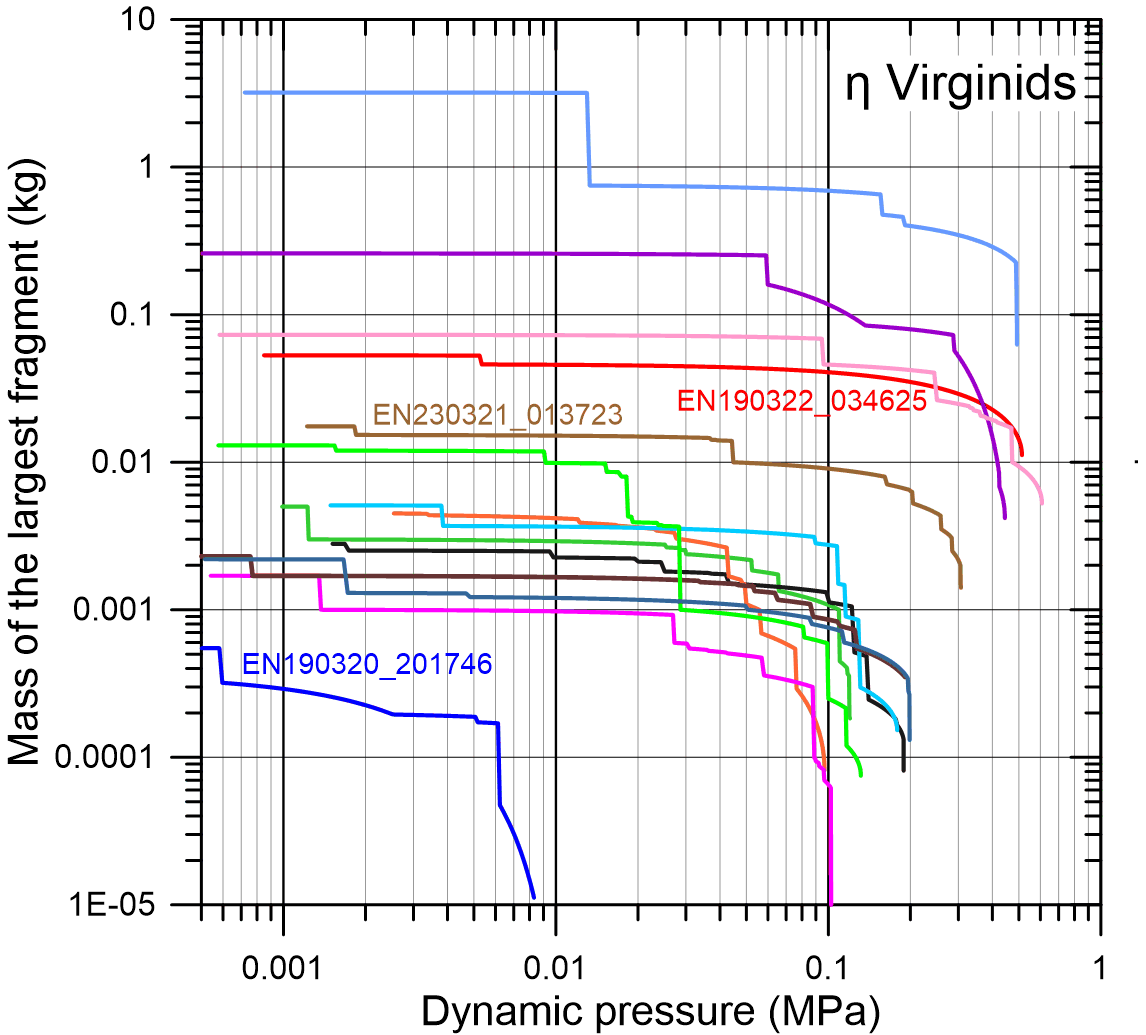}
	\caption{Mass of the largest remaining fragment as a function of increasing dynamic pressure for 14 modeled $\eta$-Virginids. 
	Colors are used to distinguish individual meteoroids. 
	The lines belonging to the two light curves shown in Fig.~\protect\ref{lc} and one very fragile meteoroid are labeled.}
	\label{press-mass}
\end{figure}

\subsection{Fragmentation strength}

Figure~\ref{press-mass} shows the mass of the largest remaining fragment during the penetration through the atmosphere as a function
of increasing dynamic pressure for all modeled $\eta$-Virginids. Sudden drops of mass are due to fragmentation, gradual decrease is due to
either erosion or ablation. The phases when the dynamic pressure decreased toward the end of the fireball are not shown.
The curves belonging to the two fireballs in Fig.~\ref{lc} are identified. The last fragmentation
of EN190322\_034625 occurred at the time when dynamic pressure reached the maximum. 

We can see that small $\eta$-Virginids reach maximum dynamic pressure 0.1 -- 0.2 MPa and large ones up to 0.5 MPa. The exception was
the smallest meteoroid EN190320\_201746 which was destroyed well below 0.01 MPa. At the same time, it was the only $\eta$-Virginid which
occurred outside the usual activity years, namely in 2020. Though its radiant and orbital elements perfectly match with $\eta$-Virginid, it was
probably a random interloper with different origin and physical properties. It will be therefore excluded from further discussion. The largest
$\eta$-Virginid EN210325\_042444 with a mass of 3.2 kg was observed during twilight, only 35 minutes before sunrise. It was a 200 km long fireball
with a slope of 15$^\circ$ to the horizontal reaching an absolute magnitude $-11$ and lasting for 7.4 seconds. The observed end height was 45.5 km. 
The dynamic data from IP cameras are good but there is no radiometric light curve.

\begin{figure}
	\centering
	\includegraphics[width=8cm]{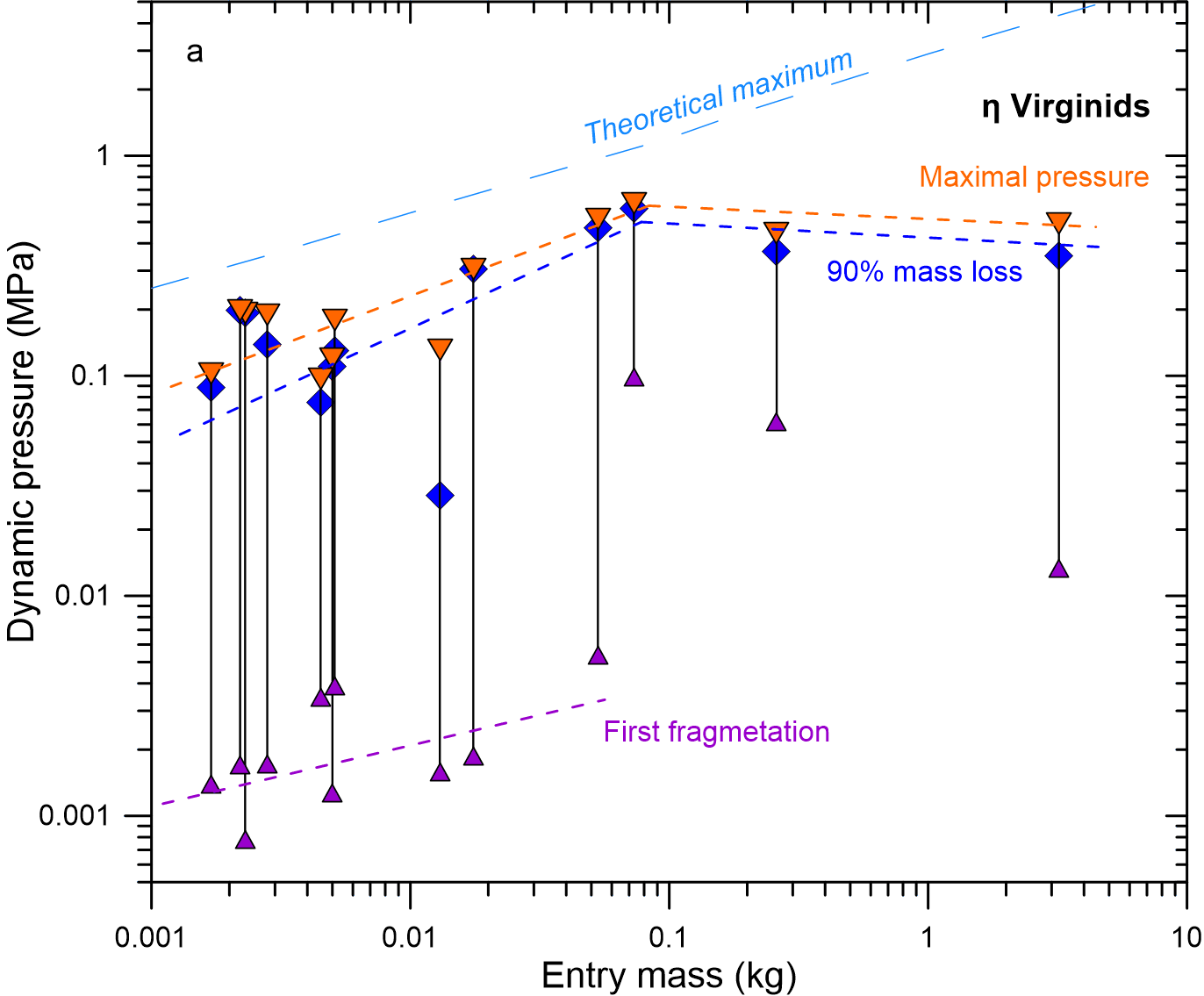}
        \includegraphics[width=8cm]{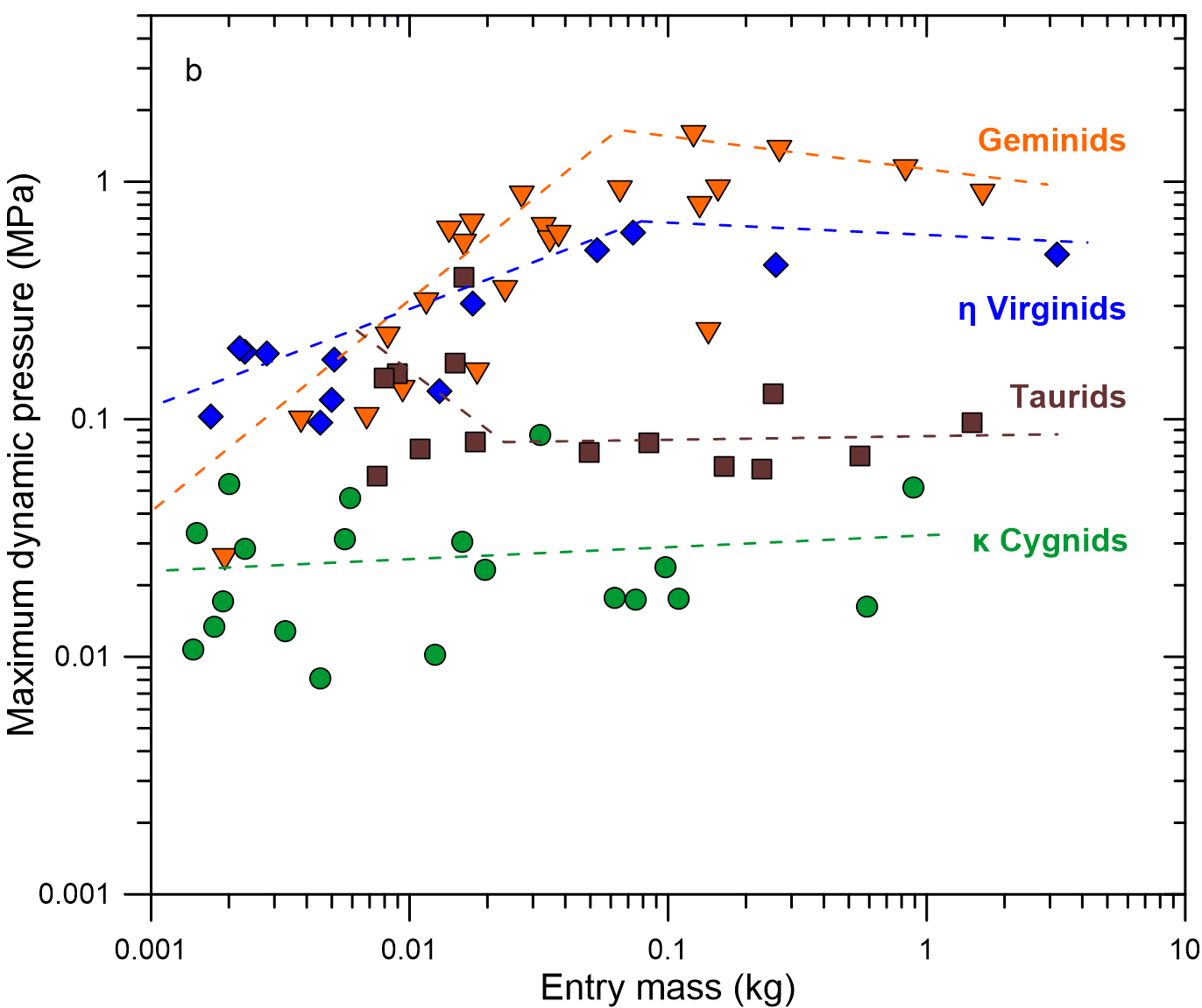}
	\caption{Selected dynamic pressures along the trajectories of all 13 modeled $\eta$-Virginid fireballs as a function of entry mass (panel a). 
	The dashed lines are general trends, not exact fits to the data. The theoretical
	maximum was computed for hypothetical non-fragmenting and high-density $\eta$-Virginids. Panel b presents a
	comparison of the observed maximum pressure
	with three other representative showers, from strong Geminids to fragile $\kappa$-Cygnids.}
	\label{pressures}
\end{figure}

In Fig.~\ref{pressures}a, dynamic pressures at the first fragmentation, at the time when the mass of the largest fragment dropped to 10\% of
the original mass, and the maximum dynamic pressure reached are plotted as a function of the original mass. The theoretical maximum dynamic
pressure which could be reached by a strong non-fragmenting $\eta$-Virginid of the given mass is also indicated. It was computed for
a meteoroid density 3500 kg m$^{-3}$, $\sigma=0.005$ s$^2$km$^{-2}$, $\Gamma A = 0.8$, and the average trajectory slope (35$^\circ$)
and entry speed (29 km s$^{-1}$) in our sample.

As we can see in Fig.~\ref{pressures}a,
the first but generally minor fragmentation started in small ($<0.02$ kg) meteoroids at pressures below 5 kPa. Some of the larger meteoroids avoided
any low-pressure disruption. With one exception, the 90\% mass loss occurred just before the maximum pressure was reached. As for the maximum pressure,
medium sized
meteoroids with mass just below 0.1 kg proved to be the strongest, reaching the pressures of 0.5 MPa, which was close to the theoretical limit
for their mass. They lost most mass by regular ablation with fragmentation playing a lower role. Larger meteoroids 
lost most mass in the form of eroding fragments and did not approach the theoretical maximum. It seems that 0.5 MPa is the material limit of $\eta$-Virginids.

The comparison of maximum pressures with those observed in three other meteor showers is presented in Fig.~\ref{pressures}b. The Taurid
data were taken from \citet{Tauridphys} and Geminid data from \citet{Henych2024}. The identification of $\kappa$-Cygnid fireballs was recently
done by \citet{Bor_KCG}. The details of their fragmentation modeling, done with the same methods as here, will be published elsewhere.
They proved to be fragile bodies, obviously of cometary origin, reaching maximal pressures of only 8--80 kPa. Taurids are also of cometary origin
but are stronger, typically 50--150 kPa. Some small Taurids of masses of about 10 grams are comparable with $\eta$-Virginids and Geminids
of similar masses. Meteoroids of the Geminids and $\eta$-Virginids behave similarly. The strength increases with mass up to about 0.1 kg, where it reaches the maximum
which does not increase further with mass. The maximum strength of Geminid meteoroids is higher than of $\eta$-Virginids, about 1.5 MPa, although there is no difference
for small meteoroids. At few grams, $\eta$-Virginids may be even stronger than Geminids.

\subsection{Bulk density}

Another important quantity is the meteoroid bulk density. It is generally difficult to determine from meteor data. In our modeling, we were able to
obtain reasonable values of meteoroid density combining both photometry and dynamics. A lower density means a larger cross section, which produces
higher luminosity and larger deceleration. The luminosity can be used when the meteoroid is a single body -- either before the first fragmentation or
after the end of the initial erosion. Deceleration was not present at the beginning of the fireballs because of low atmospheric density. It
started in the middle part. Assuming that the main fragment was a single body, deceleration could be used to determine the meteoroid density. 
In fact, the same deceleration could be alternatively produced by a body of higher density disrupted into several pieces. Deceleration therefore effectively
provides a lower limit of density. Moreover, both luminosity and deceleration are affected by the values of parameters such as meteoroid shape 
and ablation coefficient. Nevertheless, obtaining a consistent solution for both photometry and deceleration provides some confidence that
all parameters have been set reasonably well. The uncertainties of densities, assuming that other parameters are roughly correct, are about 10\%.

\begin{figure}
	\centering
	\includegraphics[width=10cm]{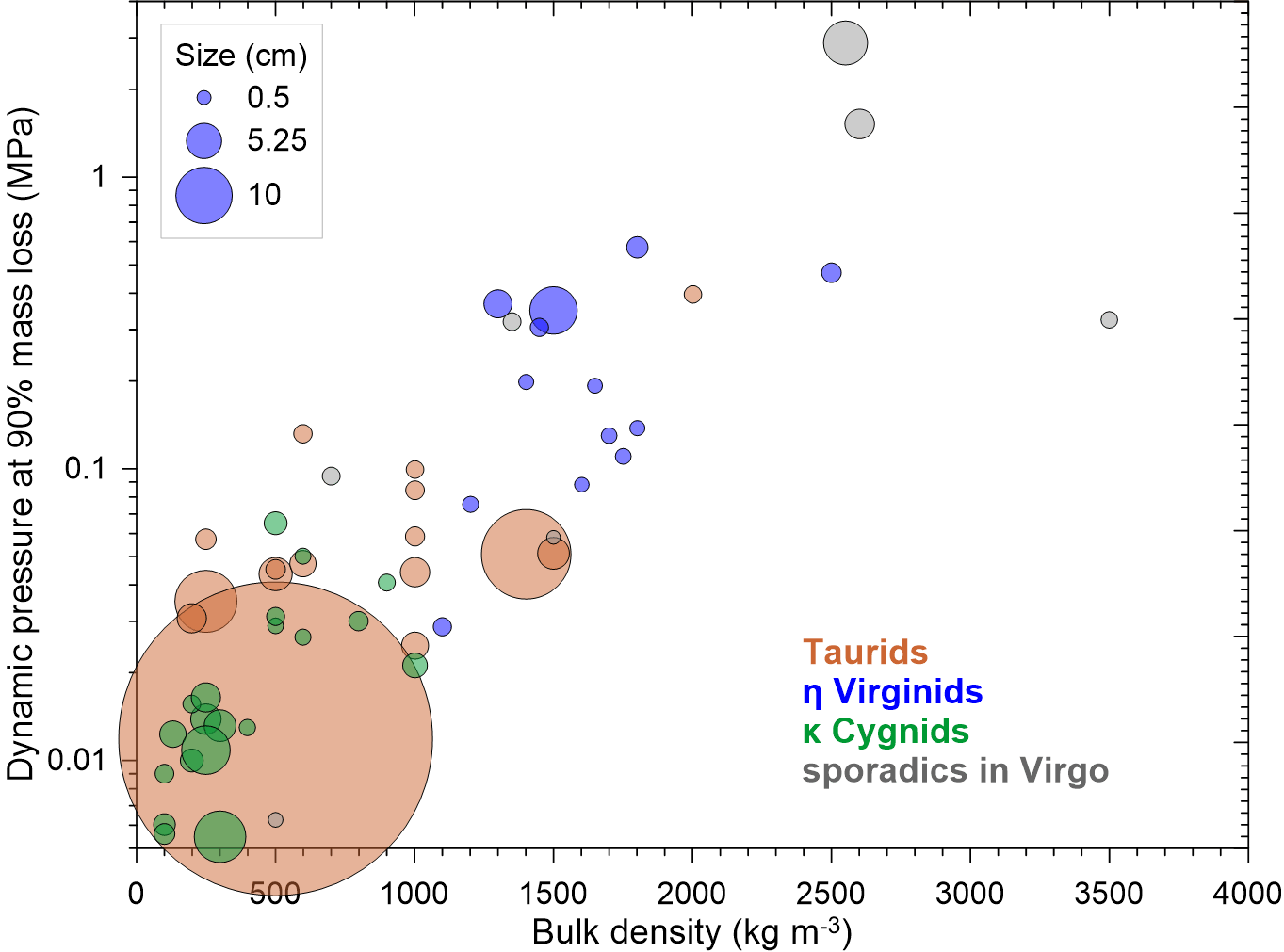}
	\caption{The relation of meteoroid bulk density, size, and mechanical strength represented by the dynamic pressure at 90\% mass loss
	for three meteor showers and some sporadic fireballs with radiants in Virgo. Symbol size is proportional to meteoroid size. Different showers
	are distinguished by different colors.}
	\label{density}
\end{figure}

The resulting densities in relation with the dynamic pressure at 90\% mass loss and meteoroid size are shown in Fig.~\ref{density}. For
comparison, Taurids, $\kappa$-Cygnids, and some of the meteoroids initially considered to be $\eta$-Virginids but later rejected are included
as well. The 90\% mass loss was taken instead of maximal pressure in order to better represent the meteoroid as a whole. The difference
was significant for some large Taurids containing small stronger inclusions. The size (diameter) was computed from the determined mass 
and density
 assuming spherical shape.

We can see that many $\kappa$-Cygnids had very low densities between 100 -- 500 kg m$^{-3}$. Some stronger $\kappa$-Cygnids
had larger densities but no more than 1000 kg m$^{-3}$. Taurids had densities 200 -- 1500  kg m$^{-3}$, in one case even 2000
kg m$^{-3}$. The typical density of $\eta$-Virginids was about 1500 kg m$^{-3}$. None had less than 1000 kg m$^{-3}$ and one
reached 2500 kg m$^{-3}$. We also tried to fit light curves and decelerations of the video observed $\eta$-Virginids using
the erosion model for small meteoroids of \citet{Draco07}. The resulting densities were about 2000  kg m$^{-3}$ or larger.
The seven sporadic meteoroids had very wide range of densities and strength confirming that they were of
various origin. Two of them were compatible in material properties with $\eta$-Virginids but very likely by chance. One of them had radiant in region B
and the other in region C.

This analysis shows that $\kappa$-Cygnids represent a very weak material, undoubtedly of cometary origin. The parent body is not known
but the orbit is of a short-period comet type. The Taurids originate from comet 2P/Encke and represent a stronger cometary material. The main
difference with $\kappa$-Cygnids is that Taurids contain small inclusions of higher strength and density. The properties of $\eta$-Virginids are different. 
The stream does not contain low-density fragile bodies. 
On the other hand, the densities and strengths do not reach the values typical  for ordinary chondrites and their precursor meteoroids. 
But they are compatible with carbonaceous chondrites containing some porosity. For example, the Winchcombe CM2 meteoroid had the strength
0.6 MPa \citep{Winchcombe2} and the recovered meteorites had the bulk density 2090 kg m$^{-3}$ \citep{Winchcombe1}. The original 
meteoroid could have a larger porosity and lower density.
Since carbonaceous chondrites originate from C-type asteroids such as Bennu \citep{Bennu} and Ryugu \citep{Ryugu}, $\eta$-Virginids are probably
of asteroidal origin. The orbits of fireballs are compatible with that since they have aphelia well inside the Jupiter orbit (Fig.~\ref{orbits}) and
below the limiting value of 4.9 AU of \citet{Bor22b}, despite the Tisserand parameter being about 2.9. 
In many aspects, $\eta$-Virginids are physically similar to Geminids, though
their strengths is somewhat lower (Fig.~\ref{pressures}b) and densities are also lower than that determined by various authors for Geminids
\citep{Babadzhanov2009, IAUS263, Henych-inpress}.

\section{Possible parent bodies}

To investigate whether the meteoroid stream is associated with a known asteroid, i.e.\ that the meteor shower has a potential parent body, 
we performed orbital-similarity analyses using the criteria  proposed by \citet{Southworth-Hawkins} and by \citet{Drummond1981}.  
We also applied the Nesvorn\'{y} method \citep{Nesvorn2006} as a complementary approach, 
since it weights the orbital elements differently from the classical D-criteria and is therefore sensitive to additional aspects of orbital proximity. 
In this context, the Nesvorn\'{y} method is used to rank candidate parent bodies according to how closely their orbits match 
the mean orbital elements of the $\eta$-Virginids (considering the observed dispersion among the meteors).
We compared these mean elements with those of approximately 1.4 million small bodies listed in the Minor Planet Center's catalog 
\footnote{\url{https://minorplanetcenter.net/iau/MPCORB.html}}. 

For each similarity method, we selected the 100 objects with the lowest criterion values and downloaded their orbital elements from
the JPL HORIZONS system\footnote{\url{https://ssd.jpl.nasa.gov/horizons/}}, adopting 2021 February 21 as the reference epoch.
With the orbital data of the potential parent bodies synchronized in time, we recomputed the similarity parameters.

\begin{table}[width=8cm]
\caption[Properties]{Similarity of the best five candidates to parent body by Southworth-Hawkins, Nesvorn\'{y}, and Drummond methods.}
\begin{tabular}{ll}
\toprule
asteroid designation & Southworth \& Hawkins  \\
\midrule
2003 FB5  & 0.000179\\
2010 VF      & 0.026141\\
2015 FP33 & 0.027765\\
2011 EF17 & 0.027822\\
2025 BC10 & 0.031426\\
\hline
asteroid designation& Drummond  \\
 \hline
2003 FB5  & 0.000481\\
2025 FA15 & 0.010379\\
2011 EF17 & 0.017692\\
2006 UF17 & 0.018344\\
2017 TF5  & 0.018456\\
\hline
asteroid designation& Nesvorn\'{y} \\
 \hline
2003 FB5  &   2.057 \\
2015 FP33 &  96.085 \\
2025 FA15 & 378.775 \\
2023 VO   & 445.913 \\
2011 EF17 & 738.176 \\
\bottomrule
\end{tabular}
\label{tab:simimilarity}
\end{table}

Assuming that the parent body's orbit may lie within the range of the meteoroid stream's orbital elements, we evaluated the minimum value of each similarity criterion
for each candidate within this range, rather than using only the mean orbital elements. Table~\ref{tab:simimilarity} presents the computed criteria for
the five best parent body candidates for each method. The 2003 FB5 figures as the best candidate using all three methods. This body was considered 
a possible source of $\eta$-Virginids by \citet{Jennbook2}. The similarity of this object with the meteoroid stream reaches orders of magnitude
lower than the other candidates. 2015 FP33 and 2025 FA15 are also strong candidates, reaching high similarity in two of the three methods.

\begin{figure}
	\centering
	\includegraphics[width=8cm]{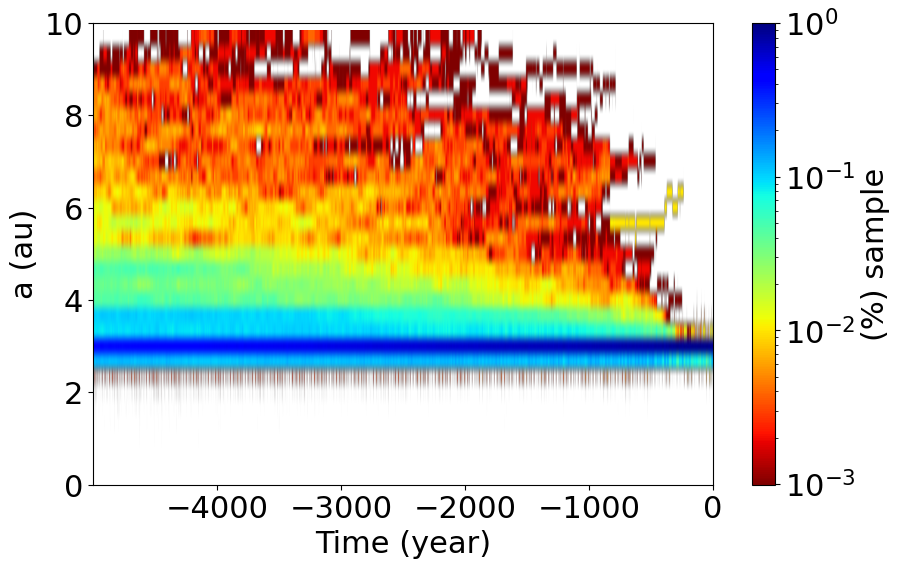}
        \includegraphics[width=8cm]{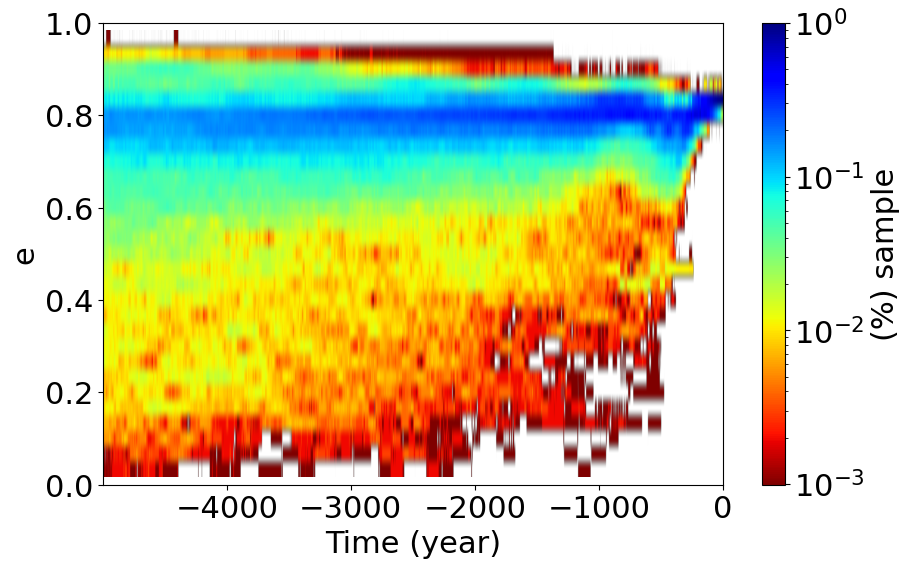}
        \includegraphics[width=8cm]{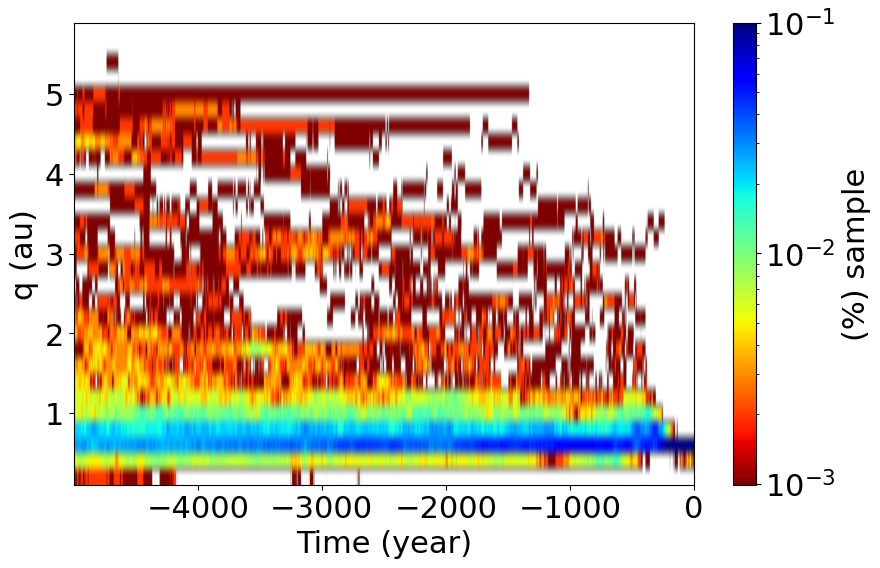}
        \includegraphics[width=8cm]{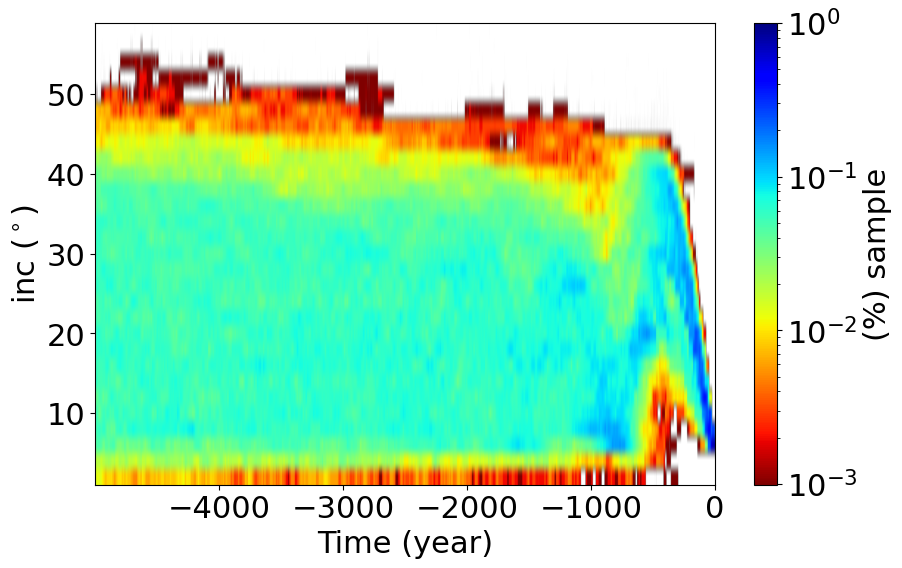}
	\includegraphics[width=8cm]{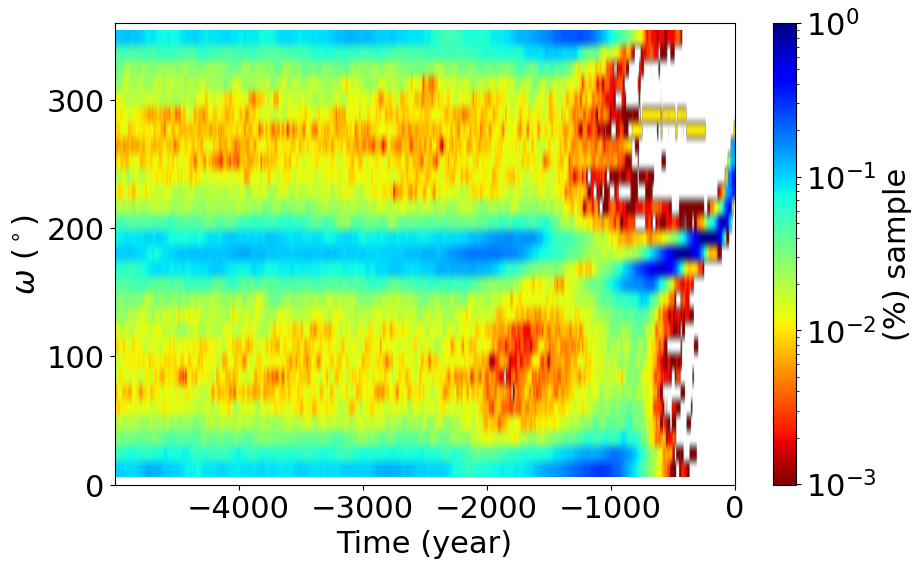}
        \includegraphics[width=8cm]{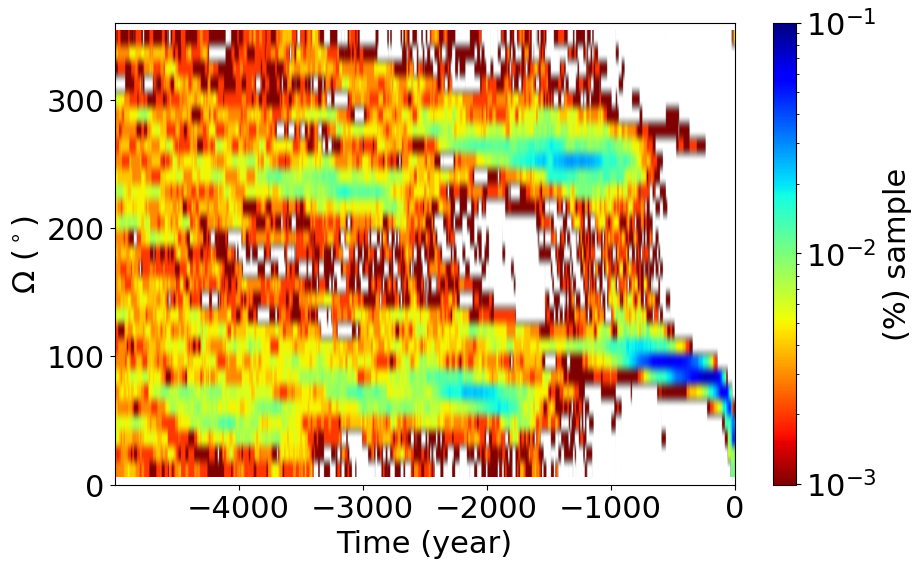}        
	\caption{Results of backward orbital integration of $\eta$-Virginids for 5000 years. The fraction of clones is color coded.
	From upper left to lower right: semimajor axis, eccentricity, perihelion distance, inclination, argument of perihelion, longitude
	of the ascending node. }
	\label{integrations}
\end{figure}

\section{Orbital evolution}

To get an insight into long-term evolution of orbits of $\eta$-Virginids, a backward orbital integration for 5,000 years was done using 
the high order integrator IAS15 in the REBOUND package \citep{rebound-ias15}.
Thousand clones with orbital elements covering the observed range of orbits of $\eta$-Virginid fireballs and their uncertainties 
were produced. The results are shown in Fig.~\ref{integrations}. The semimajor axis, eccentricity, and perihelion distance are stable. The inclination,
on the other hand, shows large variations. It can change from 5$^\circ$ to 20--40$^\circ$ within few hundreds of years. The variations seems to be periodic
but since the periods are different for different clones, the picture becomes smeared in more distant past. The longitudes of the nodes and
arguments of perihelia then cluster around two preferred values separated by 180$^\circ$. 
These variations may affect the visibility of the shower from Earth at different epochs.

It is not possible to draw any deterministic conclusions about the past or future evolution of $\eta$-Virginids from these calculations, 
since the orbits of the clones diverge drastically over time and we do
not know which clones represent real meteoroids best. 
Such an orbital dynamical behavior is expected once the meteors come from a Kirkwood gap region.
It is, nevertheless, evident that stream meteoroids can remain in the 3:1 mean motion resonance 
with Jupiter for a long time. We checked not only that the semimajor axis corresponds to the resonance but also the resonance angles. Out of 22 fireballs,
20 showed libration around the critical angle \citep[e.g.][]{Gallardo}.

\begin{figure}
	\centering
	\includegraphics[width=8cm]{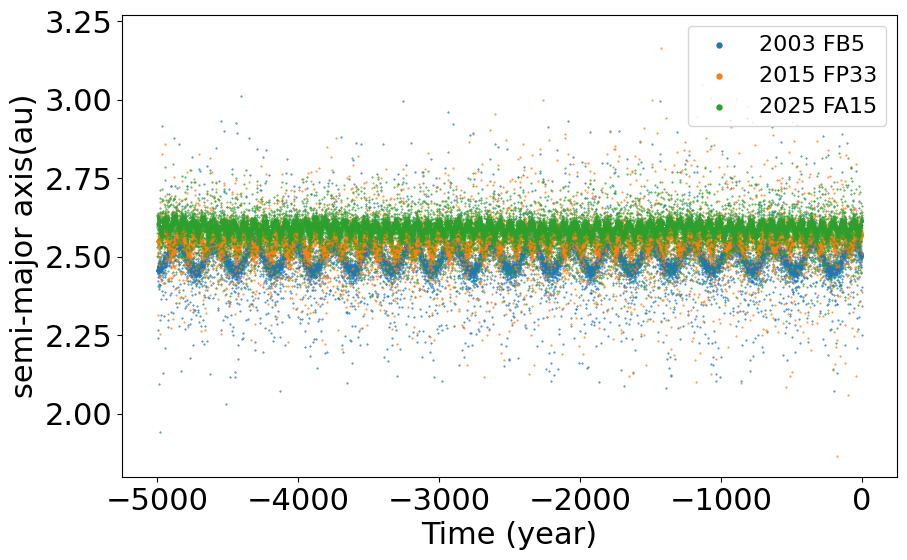}
        \includegraphics[width=8cm]{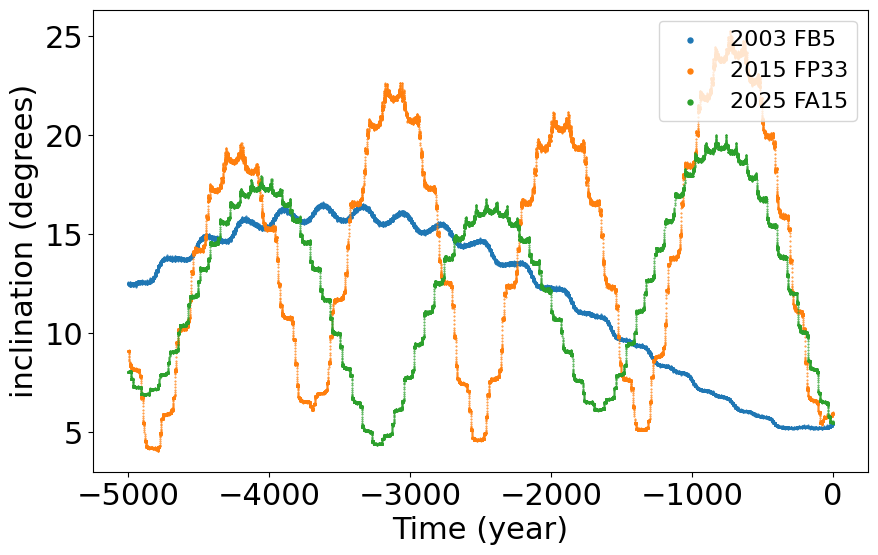}
	\caption{Evolution of semimajor axis and inclination of asteroids 2003 FB5, 2015 FP33, and 2025 FA15 from backward orbital integration.}
	\label{asteroids}
\end{figure}

Figure~\ref{asteroids} shows the evolution of semimajor axis and inclination of candidate parent asteroids 2003 FB5, 2015 FP33, and 2025 FA15.
Except small oscillations, semimajor axes remain stable. Inclinations show large variations but in case of 2003 FB5 they are much slower 
than in $\eta$-Virginids. In this respect, 2015 FP33 and 2025 FA15 are more likely to be related to $\eta$-Virginids. All three asteroids are relatively small
and were observed for only about a week. Their albedos or spectra are not known. Assuming carbonaceous composition and thus low
albedo (0.05), the diameter of 2003 FB5 is about 120 meters, of 2015 FP33 about half of that and of 2025 FA15 about 40 m. 
Their connection with $\eta$-Virginids remains uncertain. Similarly to $\eta$-Virginids, their aphelia lie well below 4.9 AU but the  
Tisserand parameters are consistent with the Jupiter-family comet region.

\section{Conclusions}

We have examined the $\eta$-Virginid meteoroid stream using data from the European Fireball Network and a double-station video experiment.
We have confirmed the four year periodicity in the activity of the meteor shower. The orbital period of four years was directly measured
for the fireballs. Somewhat surprisingly, fainter meteors showed longer periods. However, the video observations during three nights showed that
the shower is poor in faint meteors. The statistics of orbits of faint meteors are thus small at this moment. 
There is a relatively wide range of perihelion distances but only a narrow interval of inclinations. Orbital
integrations showed that the semimajor axis remains stable for thousands of years, probably thanks to the 3:1 mean motion resonance with Jupiter.
On the other hand, the inclination is expected to exhibit large periodic variations, which seems to be at odds with the small spread of inclinations
of the observed fireballs. This aspect deserves further investigation.

The physical properties of the meteoroids are different from cometary streams and are similar to the Geminids. The limiting fragmentation strength
of 0.5 MPa and typical bulk density of cm-sized meteoroids of 1500 kg m$^{-3}$ are somewhat lower than for Geminids but still suggest that the parent
body is a carbonaceous asteroid. Some small meteoroids have densities around 2500 kg m$^{-3}$. Asteroids 2003 FB5, 2015 FP33, and 2025 FA15
were identified as those with most similar orbits to $\eta$-Virginids but a genetic relationship could not be proven. In any case, $\eta$-Virginids is another stream of
asteroidal origin, in addition to the Geminids (and Daytime Sextantids related to them).
Since $\eta$-Virginids do not approach the Sun so closely as Geminids, and their perihelion distance is stable,
their high strength and density cannot be ascribed to solar proximity, strengthening the argument for asteroidal origin. 
The process of the stream formation remains, nevertheless, unknown. 

\printcredits

\section*{Declaration of competing interest}

The authors declare that they have no known competing financial interests or personal relationships that could have appeared to influence the work reported in this paper.

\section*{Acknowledgments}

This work was supported by grant no.\ 24-10143S from the Czech Science Foundation.
The operation of Slovak DAFOs was supported by Slovak Grant Agency for Science VEGA (grant
no. 2/0067/26).

\appendix
\section{Appendix}

The radiants and orbits of all individual $\eta$-Virginids are given in Table~\ref{tab:orbitall}. The formal errors can be found in the Supplementary Material. 
The masses, densities, and fragmentation
pressures derived from the fragmentation modeling are given in Table~\ref{tab:pressures} for fireballs, which could be modeled.
The zenith distance of the radiant, describing trajectory slope in the atmosphere, is also provided.

\section*{Data availability}

More detailed data about the trajectories and orbits of fireballs used in this study (not only $\eta$-Virginids) are provided in the Supplementary Material
to this article. For the modeled fireballs, the source data for modeling are also provided. They include the dynamic data (time-length-height), photographic,
and radiometric light curves. 
The file Virgo-fireball-list.xlsx contains catalog data for 118 fireballs and 8 video meteors with radiants in the Virgo region in a similar manner as
was provided for 824 fireballs in \citet{Bor22a}. Fireball classification ($PE$ and $P\!f$ numbers) cannot be used for video meteors and is not provided for them. 
Physical data such as photometric masses are given on the basis of global examination.  Fragmentation modeling
can later refine the values. Files EVI-dynamics.xlsx, EVI-photometry.xlsx, and EVI-radiometers.xlsx contain dynamic data, light curves, and radiometric light curves,
respectively, for 13 fireballs from Table~\ref{tab:pressures}. Each fireball has a separate list where data from one or more cameras are given.

\bibliographystyle{cas-model2-names}

\bibliography{evi_revised}

\begin{table*}[width=15.5cm]
\caption{Radiants and orbits of individual $\eta$-Virginids (J2000.0). The given quantities are fireball or meteor code, solar longitude, geocentric
radiant and velocity, semimajor axis, eccentricity, perihelion distance, inclination, argument of perihelion, and longitude of ascending node. }
\label{tab:orbitall}
\begin{tabular}{lllllllllll}
\toprule
Code & $\lambda_\odot$ & $\alpha_{\rm g}$ & $\delta_{\rm g}$ & $v_{\rm g}$ & $a$ & $e$ & $q$ & $i$ & $\omega$ & $\Omega$ \\
\midrule
EN140317\_194345 & 354.1817  &  181.66 &   5.31 & 26.51 & 2.507 & 0.812 &  0.470  & 5.30  & 280.13  & 354.181\\
EN150317\_0158   & 354.4408  &  182.39 &   5.29 & 26.60 & 2.46  & 0.811 &  0.464  & 5.62  & 280.96  & 354.441\\
EN150317\_233402 & 355.3370  &  186.58 &   2.25 & 29.40 & 2.53  & 0.851 &  0.378  & 5.49  & 290.40  & 355.337\\
EN120318\_231735 & 352.0788  &  179.72 &   6.37 & 26.54 & 2.546 & 0.815 &  0.472  & 5.48  & 279.75  & 352.078\\
EN100321\_191736 & 350.1501  &  178.19 &   5.61 & 26.88 & 2.498 & 0.818 &  0.456  & 4.36  & 281.72  & 350.147\\
EN140321\_205321 & 354.2089  &  179.96 &   6.47 & 25.41 & 2.548 & 0.800 &  0.509  & 5.24  & 275.58  & 354.209\\
EN150321\_011020 & 354.3868  &  182.19 &   4.69 & 26.68 & 2.471 & 0.813 &  0.462  & 5.03  & 281.23  & 354.387\\
EN150321\_232628 & 355.3116  &  186.00 &   2.76 & 28.79 & 2.49  & 0.842 &  0.395  & 5.55  & 288.64  & 355.312\\
EN210321\_003342 & 360.3323  &  188.77 &   2.31 & 27.65 & 2.583 & 0.830 &  0.438  & 5.78  & 283.59  &   0.335\\
EN230321\_013723 & 362.3614  &  190.24 &   1.45 & 27.35 & 2.541 & 0.823 &  0.445  & 5.47  & 282.99  &   2.365\\
EN180322\_232513 & 358.0373  &  185.64 &   3.78 & 26.85 & 2.538 & 0.818 &  0.462  & 5.59  & 281.06  & 358.039\\
EN190322\_034625 & 358.2177  &  188.70 &   1.91 & 28.74 & 2.480 & 0.840 &  0.396  & 5.85  & 288.56  & 358.219\\
EN210322\_191719 & 360.8468  &  187.74 &   2.50 & 26.64 & 2.54  & 0.815 &  0.468  & 5.19  & 280.37  &   0.850\\
EN210322\_232919 & 361.0206  &  188.79 &   2.06 & 27.32 & 2.574 & 0.826 &  0.448  & 5.43  & 282.47  &   1.024\\
EN170325\_224228 & 357.2379  &  186.84 &   3.06 & 28.04 & 2.495 & 0.832 &  0.420  & 5.90  & 285.83  & 357.239\\
EN190325\_013833 & 358.3538  &  187.33 &   2.73 & 27.78 & 2.517 & 0.829 &  0.430  & 5.67  & 284.69  & 358.356\\
EN190325\_024116 & 358.3971  &  183.56 &   4.57 & 25.41 & 2.557 & 0.801 &  0.510  & 4.97  & 275.56  & 358.399\\
EN200325\_014301 & 359.3508  &  187.69 &   2.91 & 27.34 & 2.517 & 0.823 &  0.445  & 5.81  & 283.03  & 359.353\\
EN200325\_024052 & 359.3907  &  188.81 &   2.04 & 28.26 & 2.554 & 0.837 &  0.419  & 5.79  & 286.03  & 359.393\\
EN200325\_182414 & 360.0415  &  187.40 &   2.85 & 26.81 & 2.525 & 0.817 &  0.462  & 5.43  & 281.07  &   0.045\\
EN210325\_042444 & 360.4557  &  188.14 &   2.07 & 27.04 & 2.484 & 0.818 &  0.451  & 5.08  & 282.44  &   0.459\\
EN210325\_225915 & 361.2241  &  188.52 &   2.40 & 26.79 & 2.517 & 0.816 &  0.462  & 5.46  & 281.12  &   1.228\\
25318040 & 358.2515 & 186.72 & 3.16 & 28.35 & 2.89 & 0.850 & 0.433 & 5.89 & 283.30 & 358.253 \\
25318057 & 358.3482 & 187.85 & 2.47 & 28.81 & 2.77 & 0.851 & 0.412 & 5.94 & 286.01 & 358.350 \\
25320049 & 360.3097 & 188.72 & 2.11 & 27.64 & 2.55 & 0.829 & 0.437 & 5.58 & 283.79 & 0.313 \\
25320055 & 360.3180 & 188.67 & 2.13 & 27.99 & 2.70 & 0.839 & 0.434 & 5.65 & 283.66 & 0.321 \\
\bottomrule
\end{tabular}
\end{table*}

\begin{table*}[width=15.5cm]
\caption{Physical parameters of modeled $\eta$-Virginid fireballs. The given quantities are fireball code, initial mass, density, zenith distance of the
radiant, dynamic pressure at the first fragmentation, at the time of 90\% mass loss, and the maximum reached pressure. Mass is in grams, density
in kg m$^{-3}$, and pressures in kPa. }
\label{tab:pressures}
\begin{tabular}{lllllllllll}
\toprule
Code & $m$ & $\rho$ &  $z_{\rm R}$ & $p_1$ & $p_{90\%}$ & $p_{\rm max}$ \\
\midrule
EN140317\_194345&  4.5  & 1200&  $62^\circ$ &  3.3 &  76 &  97  \\
EN120318\_231735&  2.8  & 1800&  41 &  1.7 &  140&  190 \\
EN100321\_191736&  5.0  & 1750&  68 &  1.2 &  110&  120 \\
EN140321\_205321&  1.7  & 1600&  51 &  1.4 &  88 &  100 \\
EN150321\_011020&  5.1  & 1700&  48 &  3.8 &  130&  180 \\
EN210321\_003342&  73   & 1800&  45 &  96  &  580&  610 \\
EN230321\_013723&  17.5 & 1450&  52 &  1.8 &  300&  310 \\
EN190322\_034625&  53   & 2500&  67 &  5.2 &  470&  510 \\
EN170325\_224228&  2.3  & 1650&  47 &  0.8 &  190&  190 \\
EN190325\_013833&  13   & 1100&  52 & 1.5  &  29 &  130 \\
EN200325\_014301&  260  & 1300&  49 &   60 &  370&  440 \\
EN210325\_042444&  3200 & 1500&  75 &  13  &  350&  490 \\
EN210325\_225915&  2.2  & 1400&  45 &   1.7&  200&  200 \\
\bottomrule
\end{tabular}
\end{table*}

\end{document}